\documentclass[fleqn,12pt]{wlscirep}
\title{Universalities of thermodynamic signatures in topological phases}

\author[1]{S.N. Kempkes}
\author[1]{A. Quelle}
\author[1,*]{Cristiane Morais Smith}
\affil[1]{Institute for Theoretical Physics, Center for Extreme Matter and Emergent Phenomena\\ Utrecht University, Leuvenlaan 4, 3584 CE Utrecht, The Netherlands}
\affil[*]{c.demoraissmith@uu.nl}
\usepackage{graphicx}
\usepackage[english]{babel}
\usepackage{amsfonts}
\usepackage{amssymb}
\usepackage{amsmath}
\usepackage{bbm}
\usepackage{tikz}
\newcommand{\onlinecite}[1]{\hspace{-1 ex} \nocite{#1}\citenum{#1}} 

\begin{document}

\begin{abstract}
Topological insulators (superconductors) are materials that host symmetry-protected metallic edge states in an insulating (superconducting) bulk. Although they are well understood, a thermodynamic description of these materials remained elusive, firstly because the edges yield a non-extensive contribution to the thermodynamic potential, and secondly because topological field theories involve non-local order parameters, and cannot be captured by the Ginzburg-Landau formalism. Recently, this challenge has been overcome: by using Hill thermodynamics to describe the Bernevig-Hughes-Zhang model in two dimensions, it was shown that at the topological phase change the thermodynamic potential does not scale extensively due to boundary effects. Here, we extend this approach to different topological models in various dimensions (the Kitaev chain and Su-Schrieffer-Heeger model in one dimension, the Kane-Mele model in two dimensions and the Bernevig-Hughes-Zhang model in three dimensions) at zero temperature. Surprisingly, all models exhibit the same universal behavior in the order of the topological-phase transition, depending on the dimension. Moreover, we derive the topological phase diagram at finite temperature using this thermodynamic description, and show that it displays a good agreement with the one calculated from the Uhlmann phase. Our work reveals unexpected universalities and opens the path to a thermodynamic description of systems with a non-local order parameter.
\end{abstract}
\maketitle

\section*{Introduction}
For a long time, it was believed that the Ginzburg-Landau formalism was able to classify all different types of phase transitions~\cite{Landau-Lifshitz, Landau}. This view changed with the discovery of the quantum Hall effect \cite{Klitzing} and topological insulators~\cite{Kane}. Topological insulators (superconductors) are materials that host symmetry-protected metallic edge states in an insulating (superconducting) bulk~\cite{Hasan, Qi2}. Although topological insulators are well understood by now, a thermodynamic description of their behavior remained elusive, firstly because the edges yield a non-extensive contribution to the thermodynamic potential, and secondly because topological quantum field theories involve non-local order parameters, and hence cannot be captured by the conventional Ginzburg-Landau formalism. Recently, this challenge has been overcome: by using Hill thermodynamics~\cite{Hill} to describe the paradigmatic Bernevig-Hughes-Zhang (BHZ) model in two dimensions, it has been shown that at the topological phase change the thermodynamic potential does not scale extensively due to the boundary effects~\cite{Quelle}. Here, we extend this thermodynamic approach to different topological models in various dimensions (the Kitaev chain and Su-Schrieffer-Heeger (SSH) model in one dimension (1D), the Kane-Mele model in 2D and the BHZ model in 3D) at zero temperature. The models all belong to the topological class $Z_2$, but are characterized by different protecting symmetries. Surprisingly, we find that all these models exhibit the same universal behavior in the order of the topological-phase transition, which depends on the dimensionality of the system. Moreover, we numerically derive the topological phase diagram for the bulk of the system at finite temperatures using this thermodynamic description, and show that it displays a good agreement with the one calculated analytically from the Uhlmann phase. Our work reveals unexpected universalities and opens the path to a thermodynamic description of systems with a non-local order parameter, provided that nonlinear terms are appropriately taken into account for the thermodynamic extensive variables. \newline

Although Hill's thermodynamics was originally conceived to describe small systems, it has been successfully used to study the critical behavior in ferromagnets~\cite{Chamberlin} and to examine the glass transition in supercooled liquids~\cite{Chamberlin2} within a mean-field cluster model. Furthermore, Hill thermodynamics is useful in large-scale models, where the systems can be considered to have a finite size due to the long-range interaction of gravity~\cite{Latella}. The way to account for the finite-size of a system is to allow for non-extensive additions to the conventional extensive thermodynamic potential $\Phi_c$. Those are described by the so-called subdivision potential $\Phi_0$,  such that the total grand potential $\Phi= \Phi_c + \Phi_0$ is split into a bulk part $\Phi_c$ and a boundary part $\Phi_0$. Recently, it was shown that Hill's thermodynamics also offers a way to describe topological states of matter, and that $\Phi_0$ captures a thermodynamic description of the edge states in the topological 2D BHZ model~\cite{Quelle}, even when the edge states merge with the bulk. As a result, Hill thermodynamics is able to give a proper thermodynamic description of a topological phase transition, whereas conventional Gibbs thermodynamics is not. Therefore, this procedure allows for a new description of topological materials within the framework of thermodynamics.\newline 
  
Other known attempts to describe topological insulators at finite temperatures were made via the Topological Entanglement Entropy (TEE) and via the Uhlmann phase, an extension of the Berry phase with which also mixed quantum states can be described. Viyuela et al.~\cite{Delgado1, Delgado2} recently used the Uhlmann phase to determine the finite-$T$ topological phase diagram for 1D and 2D topological models, where $T$ denotes the temperature. Although the Uhlmann phase is not a holonomy, and therefore is not a traditional topological invariant~\cite{Budich}, it provides an important step towards understanding the topological nature of systems at finite temperature, since the topology of the bundle of density matrices is not yet fully understood. The path ordering in the Uhlmann holonomy is not necessary for the models considered in Refs.~\onlinecite{Delgado1, Delgado2}, and therefore the Uhlmann phase description yields well-defined finite-$T$ results in these specific cases. On the other hand, the entanglement entropy  $S(\rho)= -\textrm{Tr} \left[ \rho \log(\rho) \right]$ corresponding to the ground state of a connected region A separated by a smooth boundary of length $L$ is $S_\textrm{A}= -\alpha L -\gamma$ \cite{Vidal, Kitaev2}, and although $\alpha$ is a non-universal coefficient, the TEE $\gamma$ is a universal, robust property of the gapped phase and can be related to the total quantum dimension of the system at zero $T$. In principle, the TEE can also be identified at finite $T$~\cite{Castelnovo}, but only for a conformal field theory (CFT) the TEE is well-defined and exact at zero and finite $T$. It was shown that topological order is best described via a generalization of the TEE, the entanglement spectrum~\cite{Li}, which is the full spectrum of eigenvalues of the reduced density matrices and can be related to a CFT \cite{Kitaev2, Fidkowski}.\newline 

Here, we provide a thermodynamic description for topological insulators  by properly taking the edge states into account via Hill thermodynamics and showing that they exhibit universal thermodynamic features in terms of the dimensionality and the order of the topological quantum phase transition at zero $T$. The procedure to obtain the thermodynamic potential and subdivision potential used in this work and their physical interpretation is briefly discussed in the Methods. We refer the interested reader towards Ref.~\onlinecite{Hill} for a detailed description on Hill thermodynamics and to Ref.~\onlinecite{Quelle} for the interpretation of the results for a topological insulator. We apply the procedure described in the Methods on three topological models, in three different dimensions, which are well-known in the literature~\cite{Asboth, Bernevigbook, Shen}: the 1D Kitaev chain, the 2D Kane-Mele model and the 3D BHZ model. Together with the 1D SSH model, provided in the Supplementary information, and the 2D BHZ model considered by Quelle et al.~\cite{Quelle}, our study empirically unveils a universal behavior of topological systems: a phase transition of order $D$ at the edge and of order $D+1$ in the bulk, where $D$ is the spatial dimension of the model. The details of the models are irrelevant, apart from the fact that they exhibit topological phase transitions in which a single gapless Dirac fermion appears on the edge. In addition, we analyze the finite-$T$ behavior of the entropy and the heat capacity, and construct the finite-$T$ phase diagram for the Kitaev chain, where we show that our results for the bulk display a good agreement with the ones calculated via the Uhlmann phase. To confirm the validity of our finite-$T$ results, we calculated also the phase diagram of the SSH model, which is shown in the Supplementary information. Finally, we evaluate the density of states and heat capacity for the Kane-Mele model to provide inspiration for new experiments in the field.

\section*{Results}
\subsection*{The Kitaev chain}
\begin{figure*}[h!]
\begin{tikzpicture}
\node at (8,6) {\includegraphics[width=.4\textwidth]{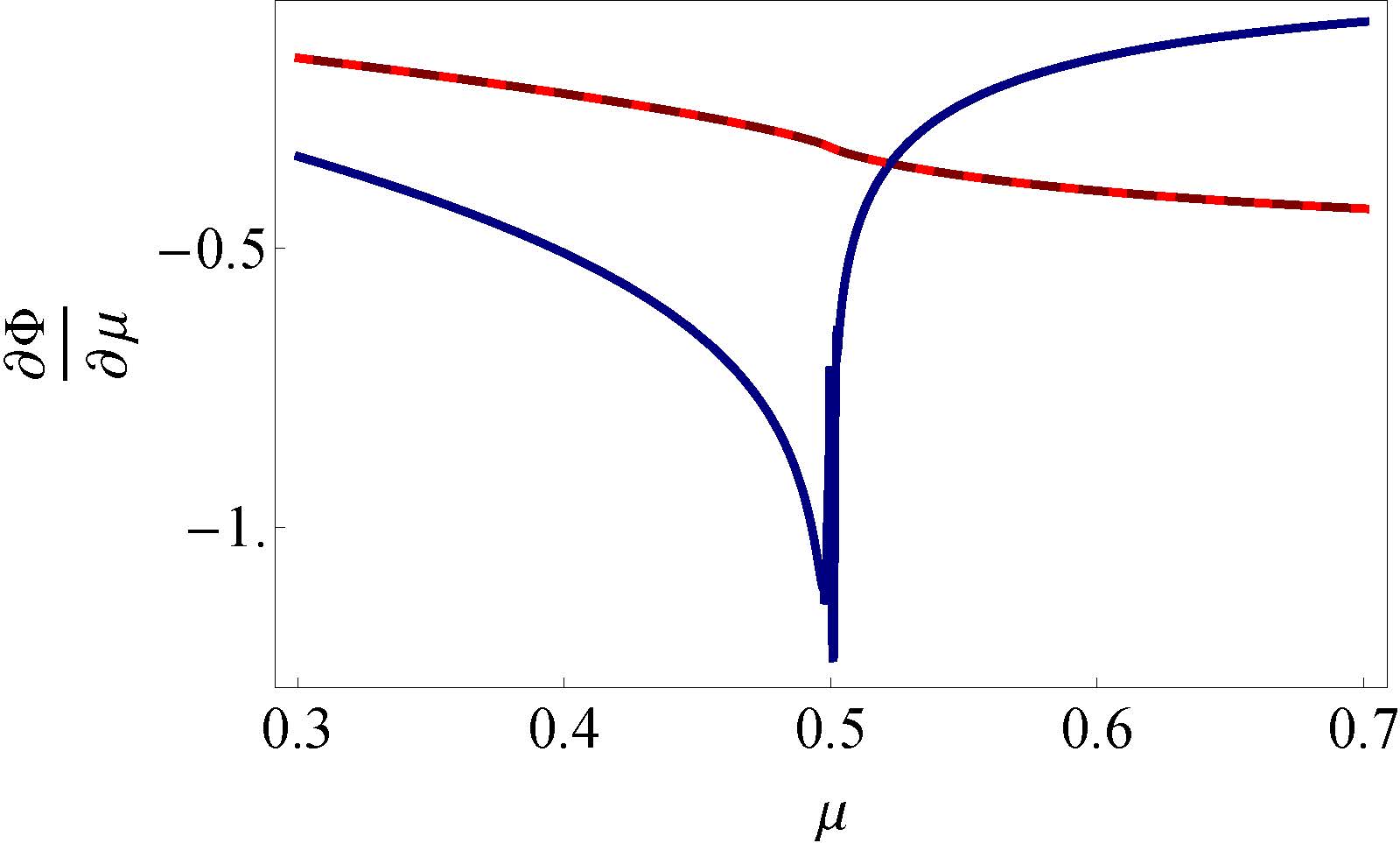}};
\node at (8,1.5) {\includegraphics[width=.4\textwidth]{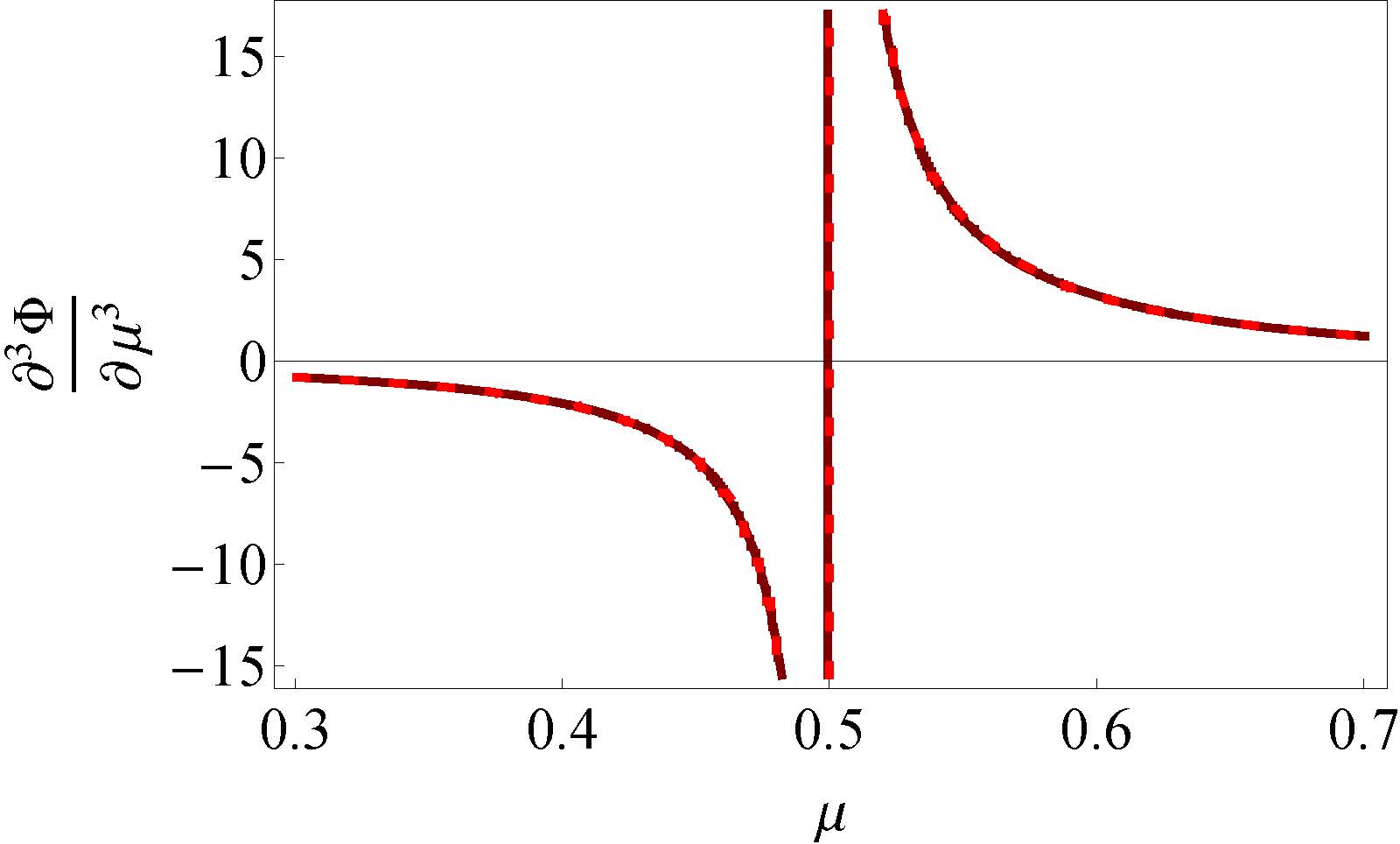}};
\node at (0,1.5) {\includegraphics[width=.4\textwidth]{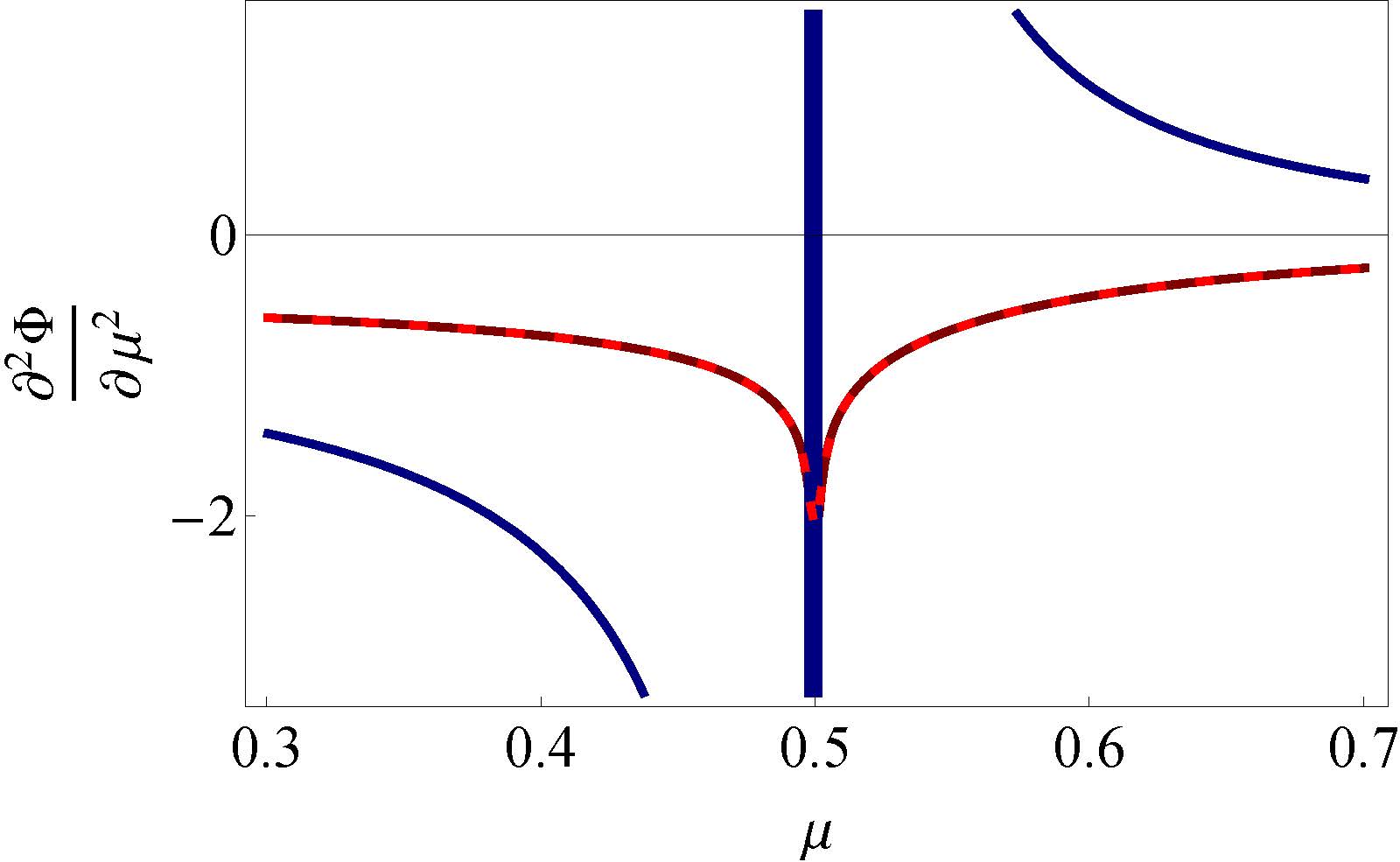}};
\node at (0,6) {\includegraphics[width=.38\textwidth]{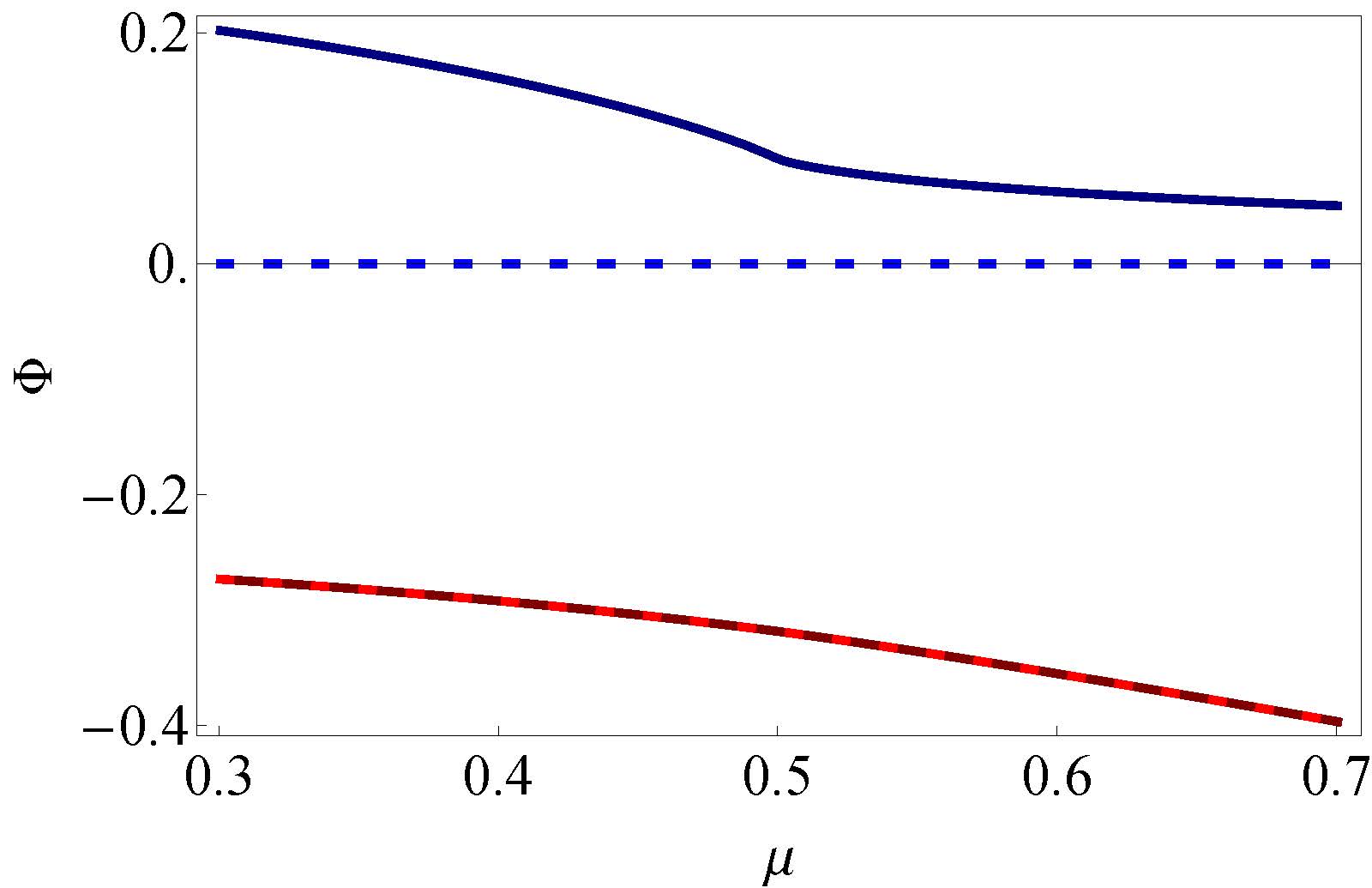}};
\node at (0,7.74) {\textrm{$\Phi_0$}};
\node at (0,5.5) {\textrm{$\Phi_c$}};
\node at (-3,8) {\textrm{\textbf{a}}};
\node at (5.3,8) {\textrm{\textbf{b}}};
\node at (-3,3.5) {\textrm{\textbf{c}}};
\node at (5.3,3.5) {\textrm{\textbf{d}}};
\node at (2,6.2) {\includegraphics[width=0.12\textwidth]{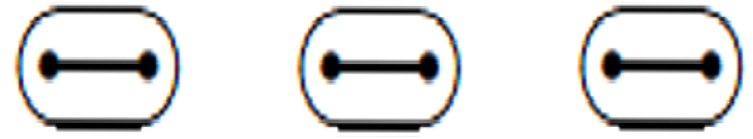}};
\node at (-1.1,6.2) {\includegraphics[width=0.12\textwidth]{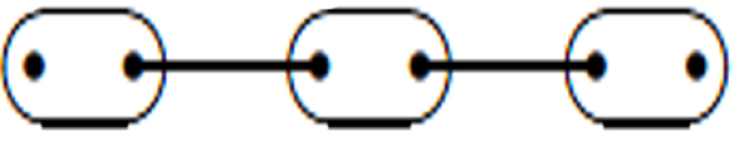}};
\node at (8,7.74) {\textrm{$\Phi_c'$}};
\node at (8,5.5) {\textrm{$\Phi_0'$}};
\node at (0,2.7) {\textrm{$\Phi_0''$}};
\node at (0,1.40) {\textrm{$\Phi_c''$}};
\node at (8,2.7) {\textrm{$\Phi_c'''$}};
\end{tikzpicture}
\caption{\textbf{Behavior of the thermodynamic potentials and their derivatives for the Kitaev chain.} \textbf{a} The conventional potential $\Phi_c$ (red) and subdivision potential $\Phi_0$ (blue) and (\textbf{b, c, d}) their derivatives with respect to $\mu$ for the Kitaev chain (solid lines) and the Kitaev ring (dashed lines), for $t=\Delta=0.25$. 
In the thermodynamic limit, the subdivision potential of the Kitaev ring indeed vanishes, whereas the conventional potential does not change. The inset in \textbf{a} shows the topological (left) and trivial (right) phase of the Kitaev chain. The topological phase hosts Majorana zero modes at the edges. The phase change occurs at $\mu=2t=0.5$. \textbf{b} The subdivision potential shows a discontinuity in the first derivative, \textbf{c} whereas the conventional potential shows a discontinuity in the second derivative. \textbf{d} The third derivative is displayed to help the visualization of the discontinuity in the second derivative. }
\label{KitaevinfiniteT}
\end{figure*}
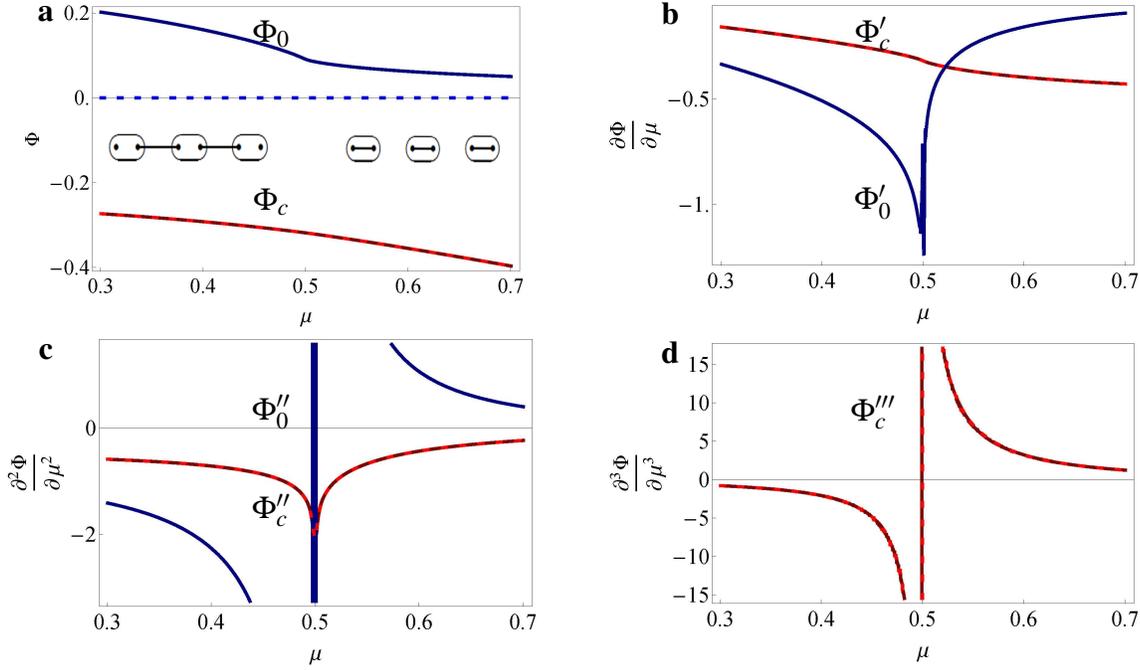

The Kitaev chain is a well-known, paradigmatic 1D model that hosts two phases, one trivial and one topological \cite{Kitaev}. The model describes the recently realized superconducting phase in indium antimonide (InSb) quantum wires contacted with a normal (Au) and a superconducting (NbTiN) electrode. The proximity to the conventional superconductor, allied to the topological features of InSb, allows for the realization of the long sought topological superconductor hosting Majorana bound states at the edge~\cite{Kouwenhoven}. The Hamiltonian of the model reads
\begin{equation}
\label{Kitaev}
H_{\text{Kitaev}} = - \mu \sum_{i=1}^N a_i^\dagger a_i - \sum_{i=1}^{N-1} \left[ t a_i^\dagger a_{i+1} - \Delta a_i^\dagger a^\dagger_{i+1} +h.c. \right],
\end{equation}
where $\mu$ denotes the chemical potential, $t \geq 0$ is the hopping parameter, $\Delta$ $\geq 0$ is the superconducting gap (pairing energy) and $a_i$ ($a_i^\dagger$) are fermionic destruction (creation) operators at position $i$, satisfying the anti-commutation relation $\{a_i,a_j^\dagger\}=\delta_{i,j}$. The phase change occurs for $|\mu| =2t$, and the system is in the topological (trivial) phase for $|\mu|<2t$ ($|\mu|>2t$). The conventional potential and the subdivision potential are calculated using statistical mechanics and by imposing an ansatz that separates the bulk and edge contributions to the thermodynamic potential (see Methods and Ref.~\onlinecite{Quelle} for a detailed description of the procedure). The results are presented in Fig.~\ref{KitaevinfiniteT}\textbf{a}, for zero $T$ and $t=\Delta =0.25$  in natural units. At $\mu=0.5$, the subdivision potential reveals a first-order phase transition at the edge (Fig.~\ref{KitaevinfiniteT}\textbf{b}), whereas the conventional potential signals a second-order phase transition in the bulk (Fig.~\ref{KitaevinfiniteT}\textbf{c}). Furthermore, when considering periodic boundary conditions (the Kitaev ring) and taking the thermodynamic limit, the edge does not contribute anymore and indeed, the subdivision potential vanishes although the conventional potential remains the same (see dashed lines in Fig.~\ref{KitaevinfiniteT}\textbf{a-d}). We further note that similar results for the phase transitions are obtained when changing phases in different ways, for example by varying the hopping parameter $t$ instead of the chemical potential $\mu$ (not shown), thus indicating that the order of the phase transition is robust to a variation of the parameters driving it. \newline

\begin{figure*}[h!]
\begin{tikzpicture}
\node at (8,6) {\includegraphics[width=.4\textwidth]{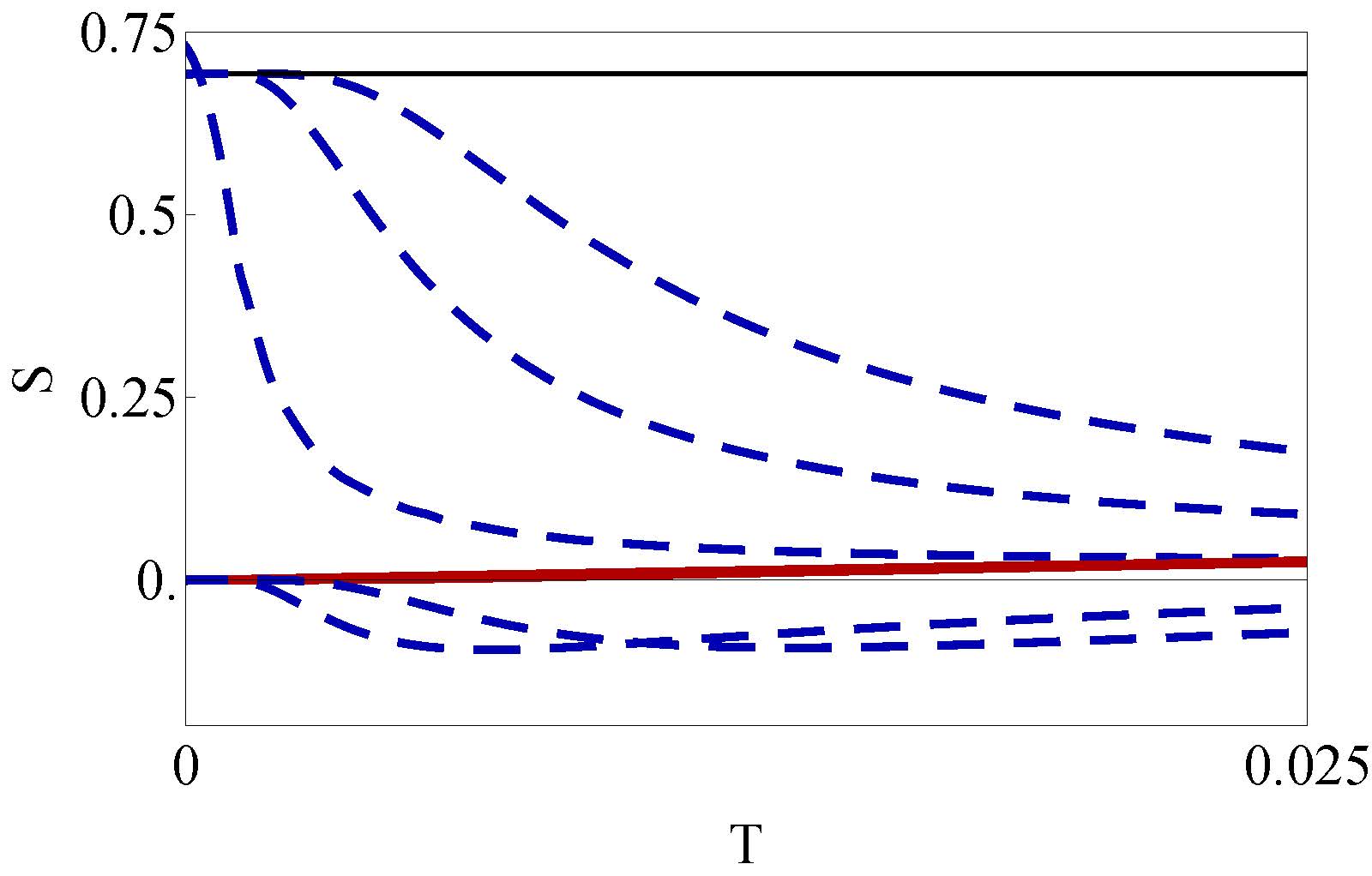}};
\node at (0,6) {\includegraphics[width=.4\textwidth]{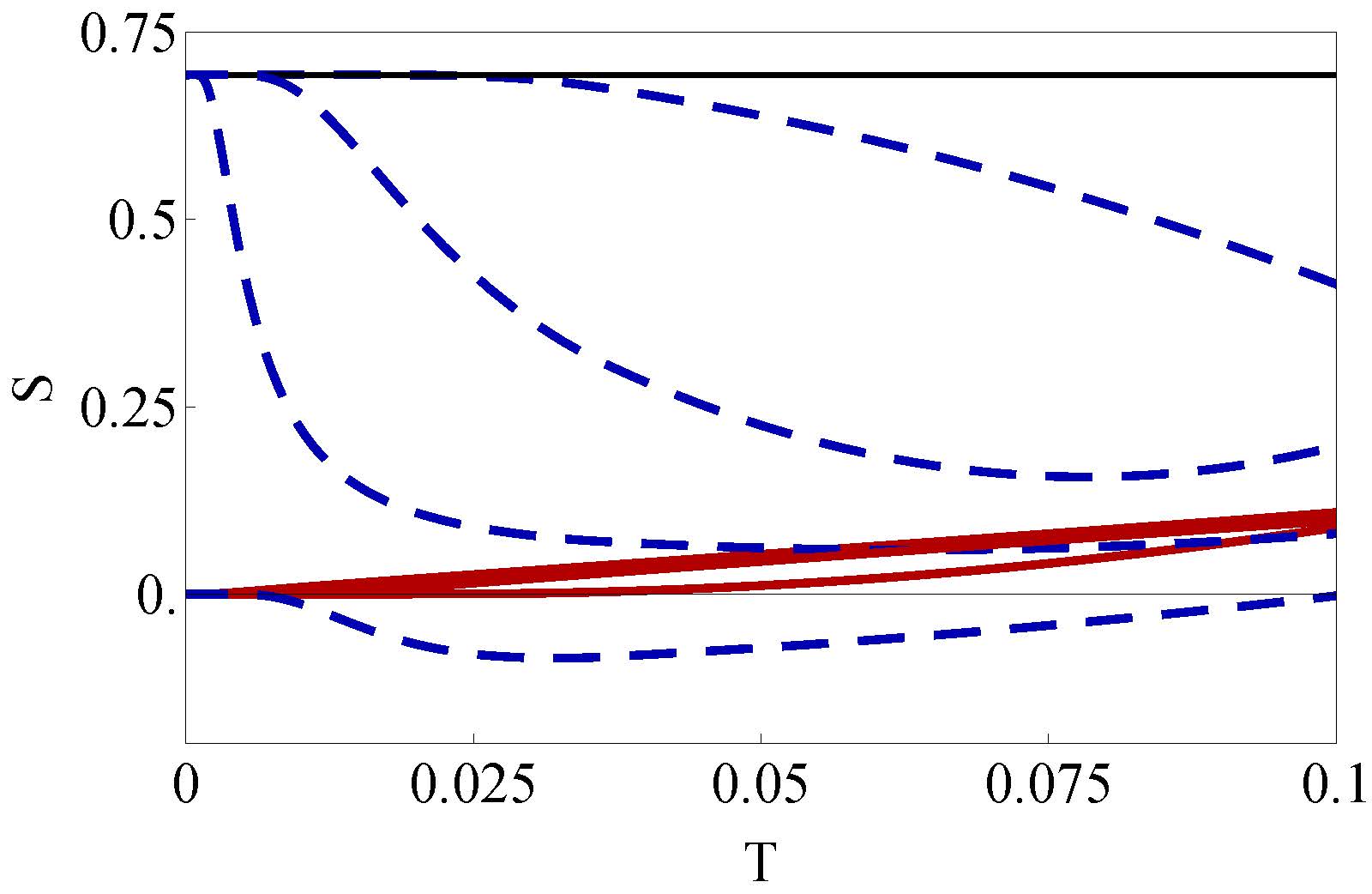}};
\node at (8,0.5) {\includegraphics[width=.4\textwidth]{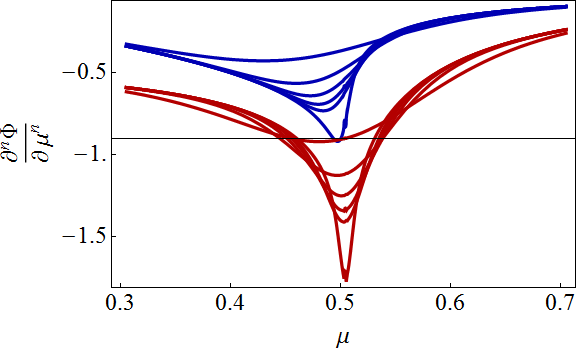}};
\node at (0,0.5) {\includegraphics[width=.4\textwidth]{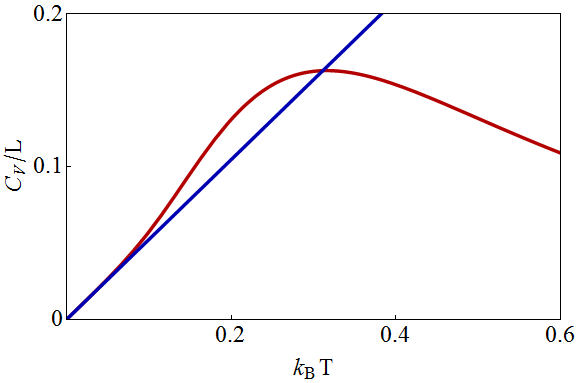}};
\node[rotate=90] at (11.8,6.5-5.5) {\textbf{\textrm{$T$} $\rightarrow$}};
\node at (-4.5,8) {\textrm{\textbf{a}}};
\node at (4.4,8) {\textrm{\textbf{b}}};
\node at (-4.5,3) {\textrm{\textbf{c}}};
\node at (4.4,3) {\textrm{\textbf{d}}};
\node at (7.5,7.7-5.5) {\textrm{$\Phi'_0$}}; 
\node at (7.5,6-5.5) {\textrm{$\Phi''_c$}}; 
\node at (2.6,7.5) {\text{\scriptsize{$\log(2)$}}}; 
\node at (1.5,6.2) {\text{\scriptsize{$\mu=0.45$}}}; 
\node at (-0.8,5.9) {\text{\scriptsize{$\mu=0.49$}}}; 
\node at (1.5,4.75) {\text{\scriptsize{$\mu=0.55$}}}; 
\node at (1.5,6.9) {\text{\scriptsize{$\mu=0.30$}}}; 
\node at (2.6+8,7.5) {\text{\scriptsize{$\log(2)$}}}; 
\node at (1.5+6.8,6.2) {\text{\scriptsize{$\mu=0.49$}}}; 
\node at (-0.8+7.79,5.9) {\text{\scriptsize{$\mu=0.50$}}}; 
\node at (1.5+8,4.75) {\text{\scriptsize{$\mu=0.52$}}}; 
\node at (1.5+5,4.75) {\text{\scriptsize{$\mu=0.51$}}}; 
\node at (1.5+6.8,7.1) {\text{\scriptsize{$\mu=0.48$}}}; 
\end{tikzpicture}
\caption{\textbf{Finite temperature effects for the Kitaev chain.} \textbf{a} Thermal entropy per lattice site in the bulk (red, solid) and at the edge (blue, dashed) for different values of $\mu=\{ 0.30, 0.45, 0.49, 0.55 \}$. The other parameters are the same as before. Upon increasing $\mu$, the entropy at the edge decreases at lower $T$ from the zero-$T$ Majorana $\log(2)$ value (black, solid), and suddenly jumps towards zero at the phase transition $\mu=0.5$ (see also \textbf{b}). The entropy in the bulk overlaps for most of the values for $\mu$ and is only visibly different for $\mu=0.30$ (lower red solid curve). 
\textbf{b} The same as \textbf{a}, but now for parameter values $\mu=\{ 0.48, 0.49, 0.50, 0.51, 0.52 \}$. Near the phase transition, the Kitaev chain is conformal, and therefore the entropy in the bulk is a linear function of $T$. \textbf{c} Low-$T$ heat capacity (in red) at the phase transition, for $t=\Delta =0.25$ (in units where $k_\text{B}/\hbar v=1$). For low $T$, we observe a linear $T$ dependence (in blue) with a slope $C_V/L k_\text{B} T=\pi/6$, which corresponds to a central charge $c=1/2$. \textbf{d} Second derivative ($n=2$ on $y$-axis) of the conventional potential $\Phi''_c$ (red) and first derivative ($n=1$ on $y$-axis) of the subdivision potential $\Phi'_0$ (blue), with respect to the chemical potential $\mu$ at different temperatures $T=$ $\{1/300, $ $1/100,$ $1/80$ ,$1/60,$ $1/40$, $1/20$$\}$. For low $T$, the phase transition is sharp, but for high $T$ it smooths out.}
\label{KitaevfiniteT}
\end{figure*}
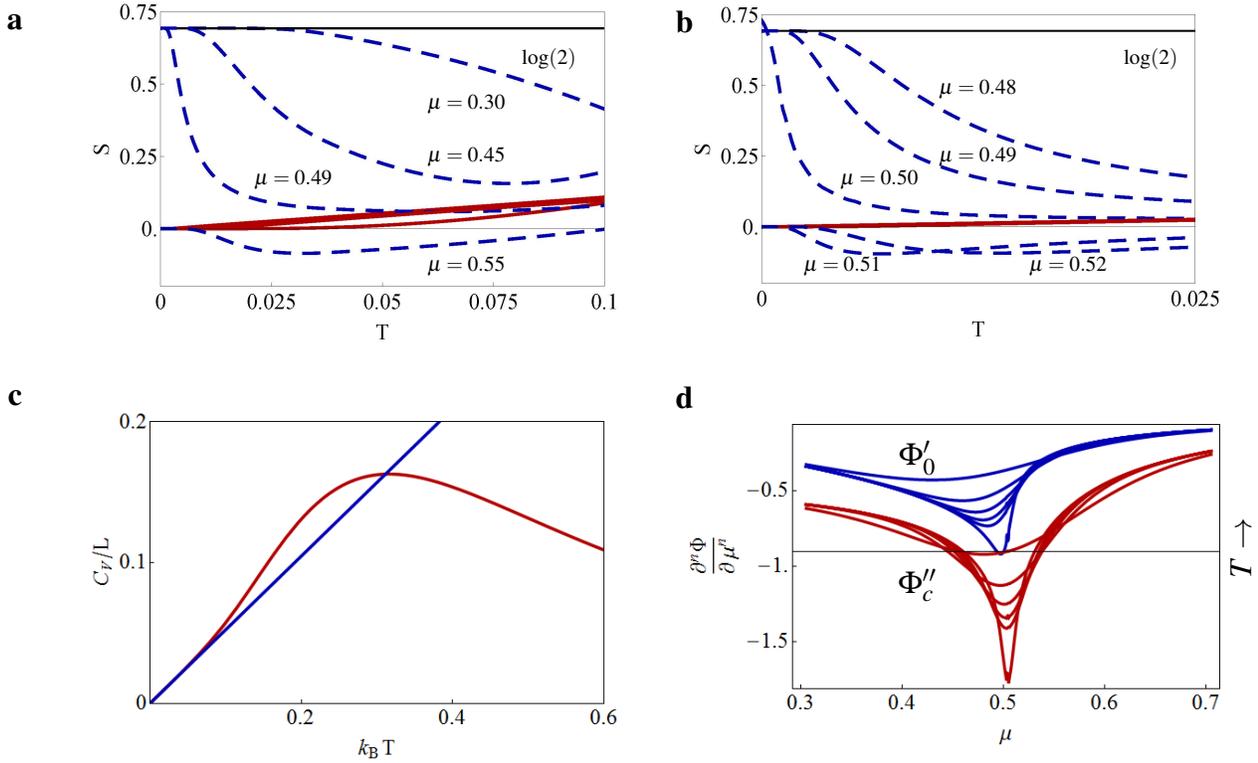

In order to verify our results, we calculate  the thermal entropy and compare the heat capacity with the one obtained via CFT.  We present the entropy $S_i=-\partial \Phi_i / \partial T$ in Fig.~\ref{KitaevfiniteT}\textbf{a}, where $\Phi_i$ denotes either $\Phi_c$ (bulk, red solid line) or $\Phi_0$ (edge, blue dashed line), for different values of $\mu$. Especially the entropy at the edge shows an interesting behavior near the phase transition, as it jumps from the initial Majorana $\log(2)$ value in the topological phase towards zero in the trivial phase (Fig.~\ref{KitaevfiniteT}\textbf{a-b}). At first glance, the negative entropy at the edge in the trivial phase seems peculiar, but it can be understood in the sense that it lowers the total entropy of the whole system. The total entropy is always positive for the considered lengths. At the phase transition, the entropy in the bulk is a linear function of $T$, as is expected for a CFT, but deviates from this linear behavior further away from it. Furthermore, the heat capacity $C_V$ of 1D quantum systems at low $T$ is proportional to the central charge c: $C_V=\pi c k_\text{B}^2 T L / ( 3 \hbar v)$ \cite{fradkin}, with $v$ the velocity of the excitations, $L$ the length of the system, $k_\text{B}$ the Boltzmann constant and $\hbar$ the reduced Planck constant. At the phase transition $\mu=2t$, the Kitaev chain is conformal, and has charge $c=1/2$. By analyzing the low-$T$ behavior of the Kitaev chain and choosing suitable parameters such that $k_\text{B}/(\hbar v)=1$, we indeed find $C_v = \pi k_\text{B} T /6 $ per unit length, see Fig.~\ref{KitaevfiniteT}\textbf{c}. \newline

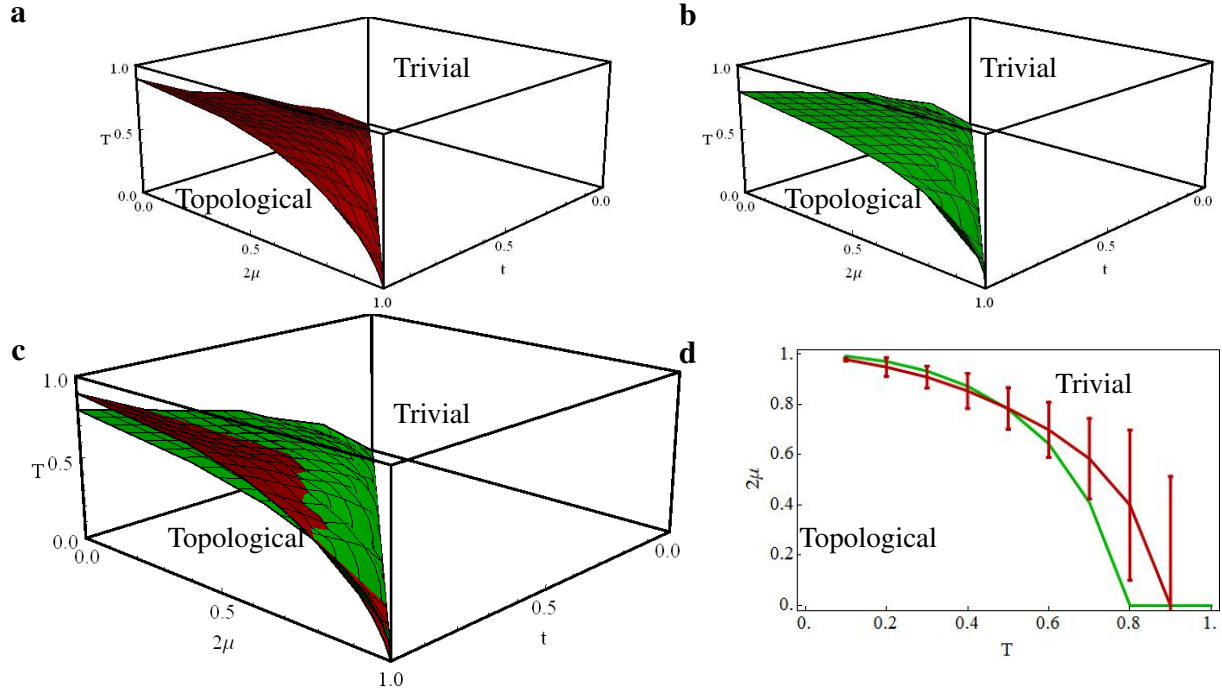
\begin{figure*}[h!]
\begin{tikzpicture}
	\node at (0,6-6) {\includegraphics[width=.52\textwidth]{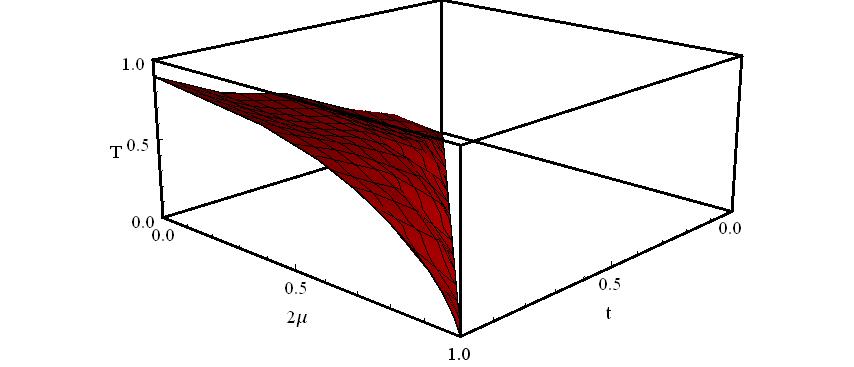}};
	\node at (8,6-6) {\includegraphics[width=.52\textwidth]{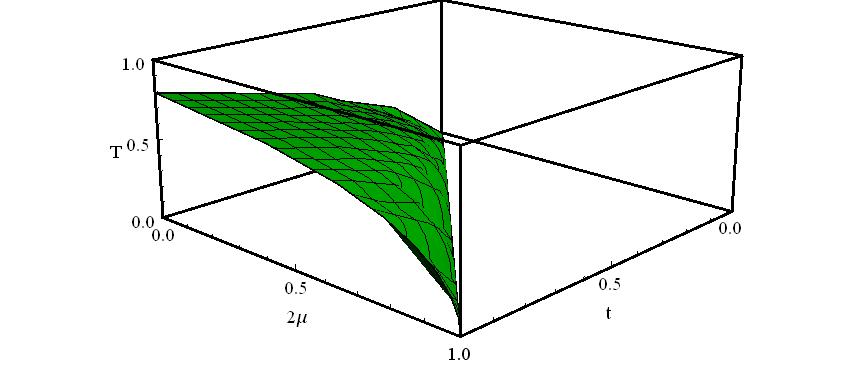}};
\node at (0,0.5-5.0) {\includegraphics[width=.55\textwidth]{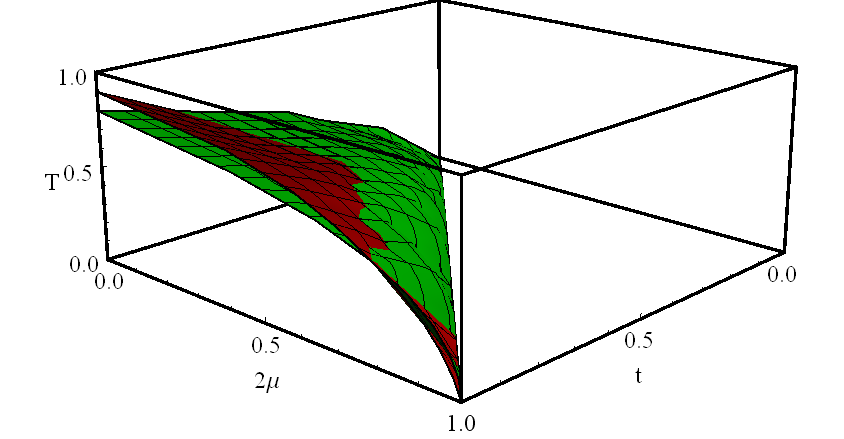}};
\node at (8.3,0.5-5.0) {\includegraphics[width=.36\textwidth]{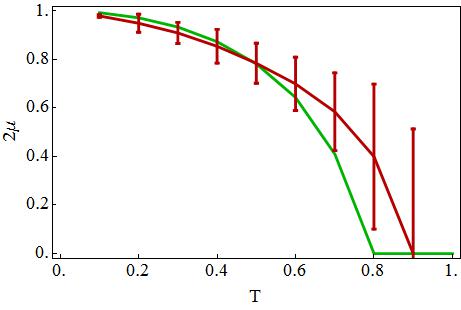}};
\node at (-4.5,3-5.5) {\textrm{\textbf{c}}};
\node at (4.4,3-5.5) {\textrm{\textbf{d}}};
\node at (-1.6,-5.0) {\small{\text{Topological}}};
\node at (1.0,-3.3) {\small{\text{Trivial}}};
\node at (-2.7+9.5,-5.0) {\small{\text{Topological}}};
\node at (1.0+8.8,-2.9) {\small{\text{Trivial}}};
\node at (-4.5,8-6) {\textrm{\textbf{a}}};
\node at (4.4,8-6) {\textrm{\textbf{b}}};
\node at (-1.5,5.52-6) {\small{\text{Topological}}};
\node at (1,7.5-6.2) {\small{\text{Trivial}}};
\node at (-2.8+9.4,5.52-6) {\small{\text{Topological}}};
\node at (2.1+6.7,7.5-6.2) {\small{\text{Trivial}}};
\end{tikzpicture}
	\caption{\textbf{Finite-$T$ phase diagram of the Kitaev chain} 
 \textbf{a} The finite-$T$ phase diagram obtained using Hill thermodynamics. Here, we use the same parameters as in Ref.~\onlinecite{Delgado1}. \textbf{b} The finite-$T$ diagram constructed  from analytical calculation of the Uhlmann phase, as shown in Ref.~\onlinecite{Delgado1}.  \textbf{c} The comparison between the two previous phase diagrams using Hill thermodynamics (red) and the Uhlmann phase (green). The red curve is below the green curve for lower values of $T$, and hence not much visible in the figure. \textbf{d} A slice of the topological phase diagram for $t=1.0$, depicting the minima of the second derivative  of the conventional potential (red) and the Uhlmann phase (green). Note that in this figure, $T$ is on the $x$-axis and $2\mu$ is on the $y$-axis. The error bars indicate a margin of error of 0.25\%  around the value of $\Phi_c ''$ corresponding to the minimum. }
	\label{kitaevphasediagram}
\end{figure*}

Now, we proceed by extending the method to finite $T$, aiming at obtaining the complete topological phase diagram. Here we focus on the Kitaev chain, but a similar analysis can be done for the other models (see Supplementary information for the finite-$T$ phase diagram of the SSH model). In Fig.~\ref{KitaevfiniteT}\textbf{d}, we display the discontinuous derivatives of the conventional potential (${\Phi_c}''$, red) and of the subdivision potential (${\Phi_0}'$, blue) for different $T$, using the same parameters as before. At high $T$, the phase transition smooths out, the discontinuity becomes less sharp, and occurs at different values of $\mu$ than for zero $T$. By considering the kink as the point that characterizes the bulk phase transition (the minima of the discontinuous derivative), it is possible to construct a phase diagram for the Kitaev chain as given in Fig.~\ref{kitaevphasediagram}\textbf{a}, where the topological (trivial) phase lies within (outside) the red coloured volume.  \newline

The topological phase diagram is compared with the results from Viyuela et al.~\cite{Delgado1}, see Fig.~\ref{kitaevphasediagram}\textbf{b}, obtained by computing the Uhlmann phase  $\Phi^\gamma_{\small{\textrm{U}}}=\arg{\textrm{Tr} \left[\rho(0) H^\gamma_{\small{\textrm{U}}}\right]}$ of a loop $\gamma$, with the Uhlmann holonomy $H_{\small{\textrm{U}}}^\gamma$ and some initial-density matrix $\rho(0)$. The Uhlmann phase is analogous to the Abelian Berry phase $\Phi^\gamma=\arg{\left<\psi(0)|\psi(T)\right>}$, where $|\psi \rangle \langle \psi|$ is a family of pure states, up to one crucial difference~\cite{Budich}: whereas the Berry phase is a U(1) holonomy and is additive in its group structure, which is essential for the construction of the invariant Chern number, the Uhlmann phase is the trace of a holonomy. This is necessary because all the characteristic classes of the bundle of density matrices vanish, hence more subtle constructions are needed to characterize the relevant topology. However, in the 1D and 2D case, this is relatively straightforward since path ordering is unnecessary~\cite{Delgado3}. \newline 

The finite-$T$ $\mu-t$ phase diagrams obtained using Hill thermodynamics to evaluate the minimum of the discontinuous derivative of the bulk potential (red) or using the Uhlmann phase (green), depicted together in Fig.~\ref{kitaevphasediagram}\textbf{c} for comparison, are actually very similar (see also Fig.~\ref{kitaevphasediagram}\textbf{d}, where the same results are shown only for a slice of the volume corresponding to $t=1.0$). The deviations are due to the fact that, at finite $T$, it is less obvious to identify the onset of the phase-transition regime, since the peak broadens as $T$ increases. We include an error bar to the bulk phase-transition points, which indicates the range of the minimum $+0.25\%$. The similarities in the results between these two fundamentally different approaches is extraordinary, and will hopefully trigger further theoretical and experimental investigations into their relation. \newline

Recently, a different topological invariant has been proposed in Ref.~\onlinecite{Budich} to describe finite-$T$  topological phase transitions, which does not shift from the zero-$T$ value as $T$ increases. We would like to point out that the inflection points of $\Phi_c'''$  always remain at $\mu=0.5$, independently of $T$, see also Fig.~\ref{KitaevinfiniteT}\textbf{d}. Although the thermodynamic approach does not contain any results on the topological structure of the bundle of density matrices, it nevertheless sheds a new light on topological materials by linking various results on topology to features of the free energy. From this perspective, it is interesting to observe that the Uhlmann phase and our numerical results capture the smoothing of the phase transition at finite $T$, following the expected finite-$T$ behavior of a quantum phase transition~\cite{Sachdev}. \newline

Now, let us investigate other topological models in higher dimensions, to try to unveil universalities in their behavior. 

\subsection*{The Kane-Mele model}
\begin{figure*} [h!]
	\begin{tikzpicture}
\node at (8,6) {\includegraphics[width=.4\textwidth]{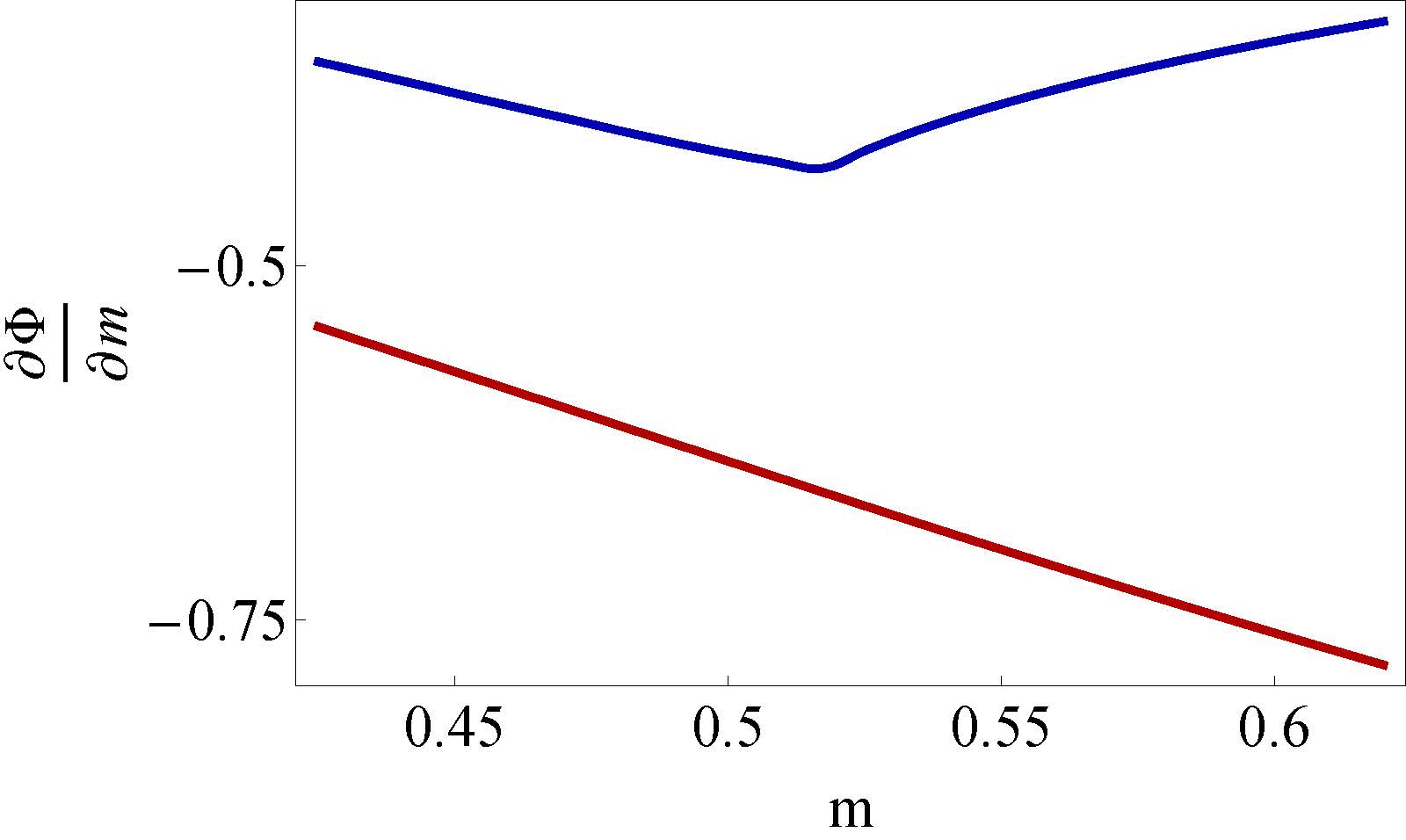}};
\node at (8,1.5) {\includegraphics[width=.4\textwidth]{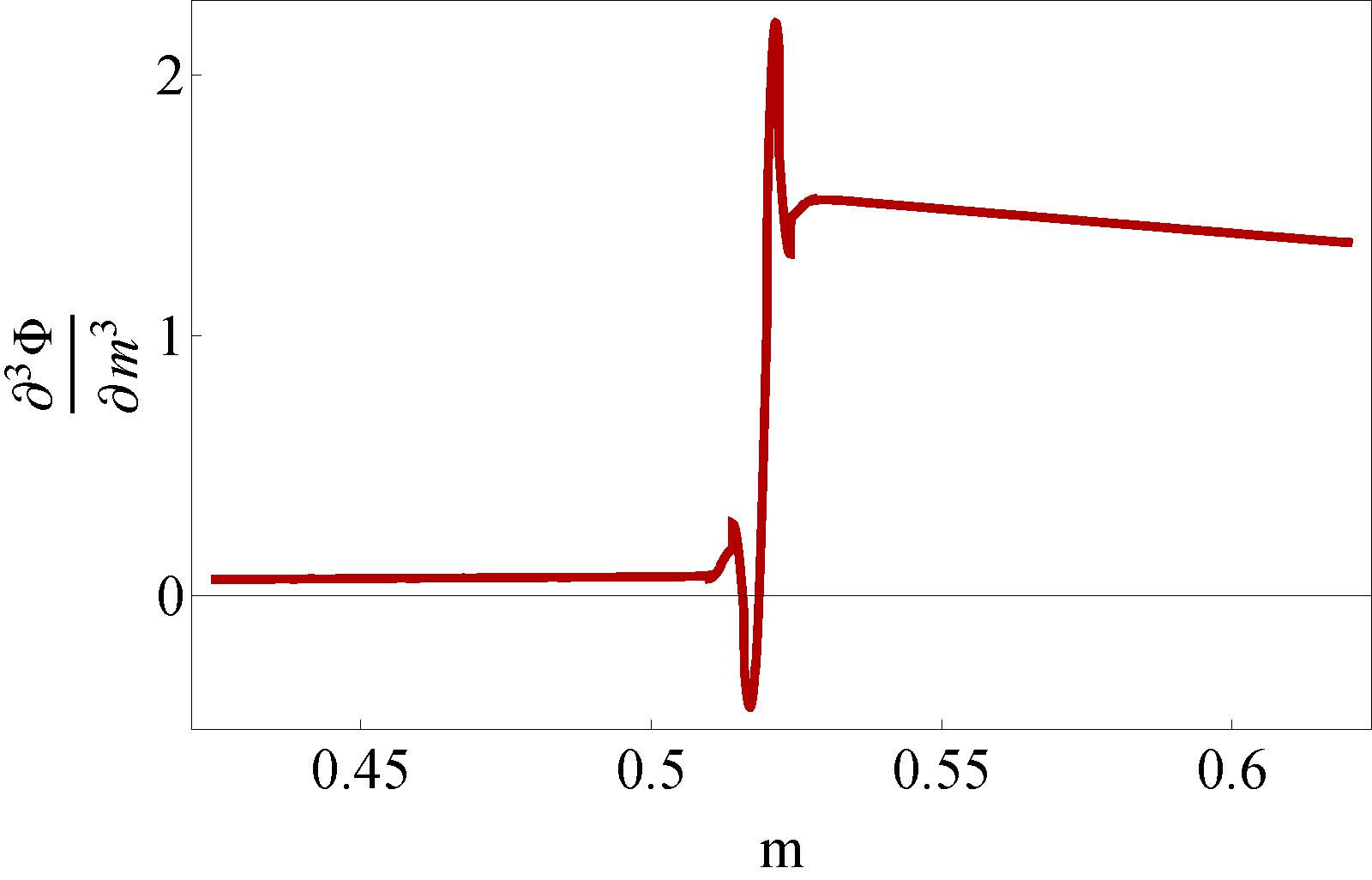}};
\node at (0,1.5) {\includegraphics[width=.4\textwidth]{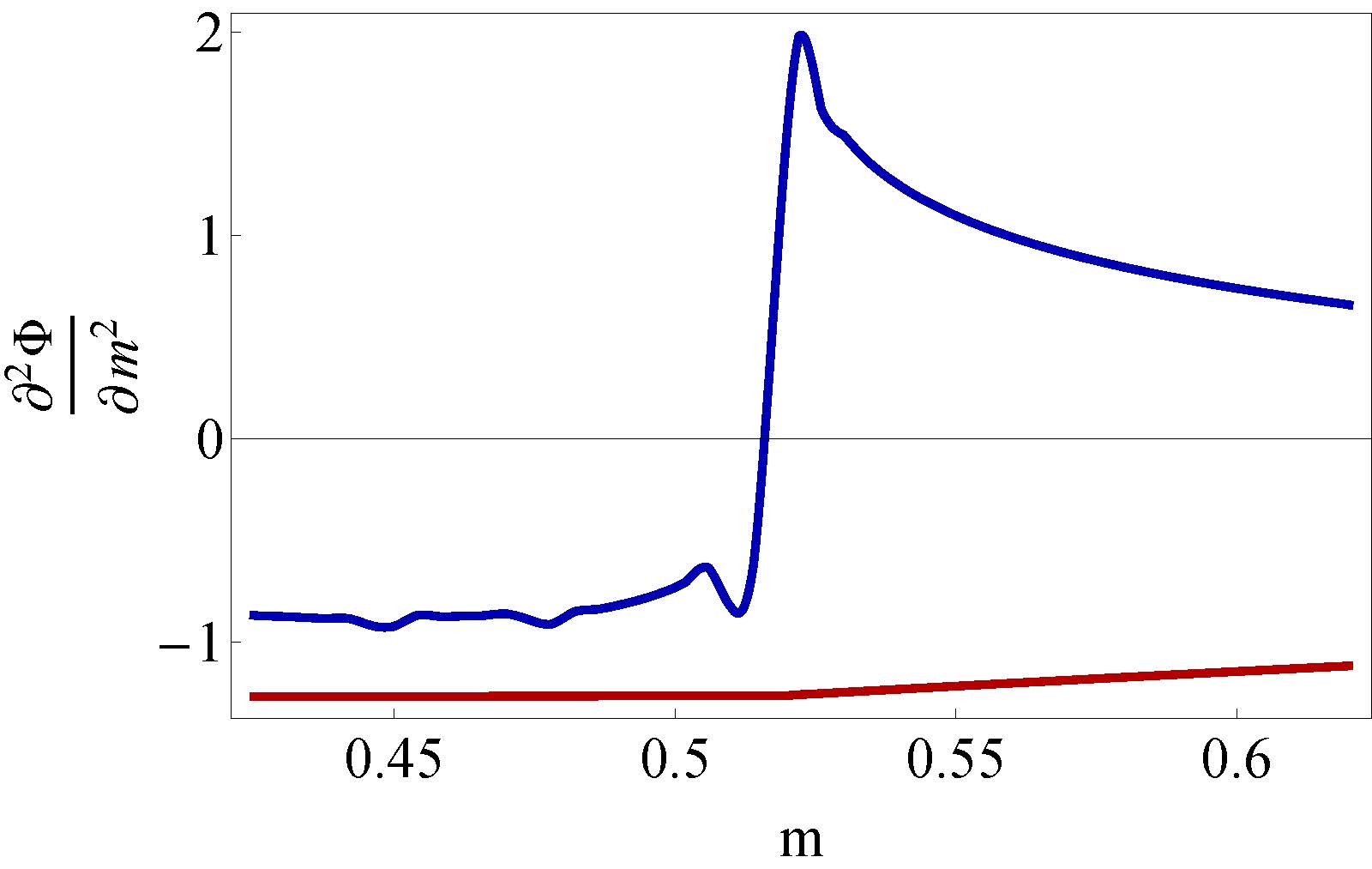}};
\node at (0,6) {\includegraphics[width=.4\textwidth]{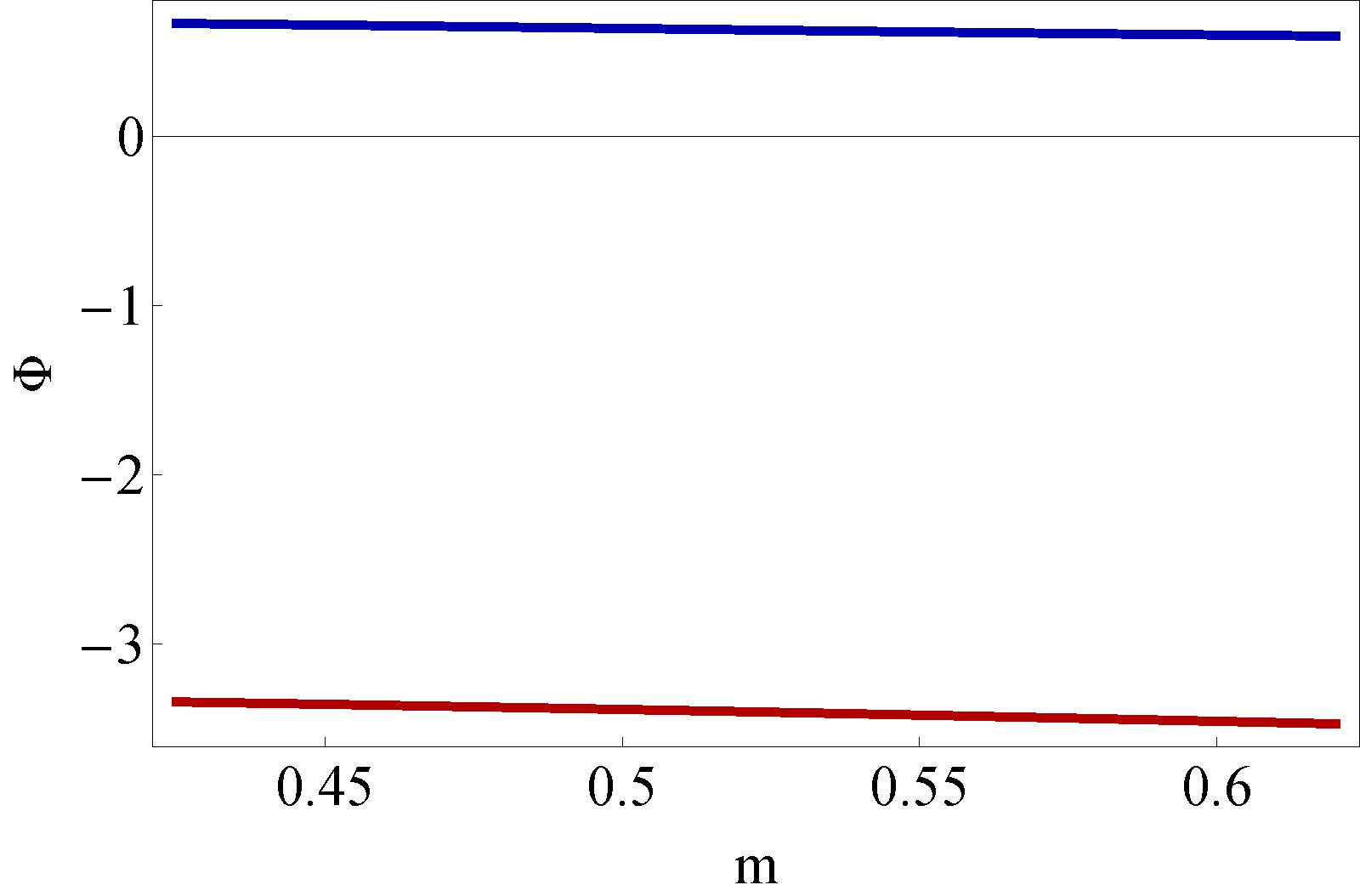}};
\node at (2,7.75) {\textrm{$\Phi_0$}};
\node at (2,5.30) {\textrm{$\Phi_c$}};
\node at (-4.5,8) {\textrm{\textbf{a}}};
	\node at (4.4,8) {\textrm{\textbf{b}}};
	\node at (-4.5,3.5) {\textrm{\textbf{c}}};
	\node at (4.4,3.5) {\textrm{\textbf{d}}};
	\node at (-0.5,6.2) {\includegraphics[width=0.12\textwidth]{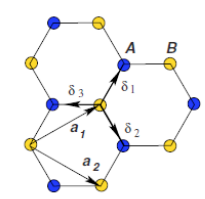}}; 
\node at (8,7.64) {\textrm{$\Phi_0'$}};
\node at (8,5.30) {\textrm{$\Phi_c'$}};
\node at (2,2.7) {\textrm{$\Phi_0''$}};
\node at (2,1) {\textrm{$\Phi_c''$}};
\node at (8,2.7) {\textrm{$\Phi_c'''$}};
	\end{tikzpicture}
\caption{\textbf{Behavior of the thermodynamic potentials and their derivatives for the Kane-Mele model.} \textbf{a} The conventional potential $\Phi_c$ (red) and the subdivision potential $\Phi_0$ (blue) and (\textbf{b, c, d}) their derivatives with respect to $m$ for the Kane-Mele model with $t=2 t_2=0.2$. The inset in \textbf{a} shows the honeycomb lattice of graphene, where the $A$ and $B$ sublattices are depicted in blue and yellow respectively. The phase change occurs at $m=0.52$. The subdivision potential shows a discontinuity in the second derivative (\textbf{c}), whereas the conventional potential shows a discontinuity in the third derivative (\textbf{d}).}
\label{Kanemele}
\end{figure*}
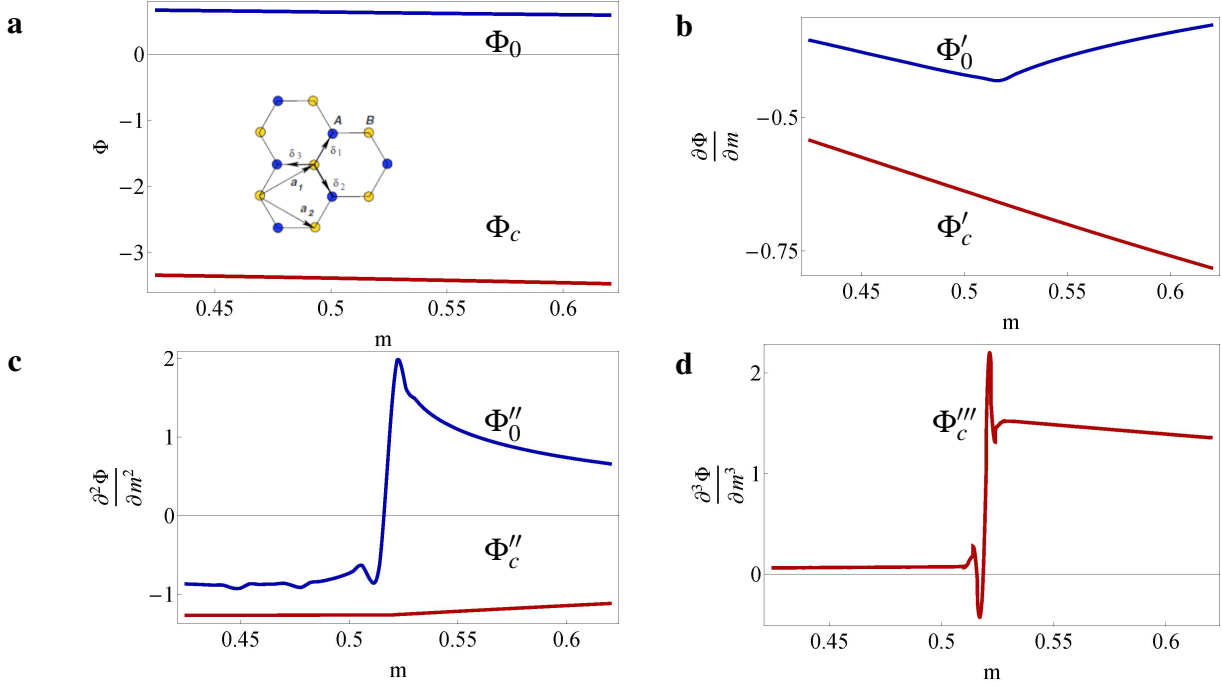

The Kane-Mele model is a paradigmatic 2D model to describe graphene, accounting for the spin-orbit coupling (SOC)~\cite{KaneMele}. By using this model,  Kane and Mele showed that in the presence of a strong intrinsic SOC there are counter-propagating quantized spin currents at the edges of the material (quantum spin Hall effect). The Hamiltonian of the model reads 
\begin{equation}
H_{\text{KM}}= \sum_{\left< ij \right> \alpha} t c_{i \alpha}^\dagger c_{j \alpha} +m \sum_i \epsilon_i c_i^\dagger c_i +   \sum_{\left<\left< ij \right>\right> \alpha \beta} i t_2 \nu_{ij} s_{\alpha \beta}^z c_{a \alpha}^\dagger c_{j \beta} ,
\end{equation} 
where the first term describes the nearest-neighbor hopping with amplitude $t$, the second describes a staggered on-site potential $m$ with opposite signs $\epsilon_i = \pm1$ for sublattices $A$ and $B$, and the third describes the intrinsic spin-orbit coupling (SOC), with $t_2$ the SOC parameter, $s^z$ a Pauli matrix representing the electron\rq{s} spin and $\nu_{ij}=-\nu_{ji} = \pm 1$ describes clock- or counter-clockwise electron hopping. For simplicity, we omit a possible Rashba term. In Fig.~\ref{Kanemele}\textbf{a-d}, we display the results for the conventional potential and its derivatives for $t=2 t_2=0.2$ and different values of $m$. We observe a second-order phase transition at the edge (Fig.~\ref{Kanemele}\textbf{c}) and a third-order one in the bulk (Fig.~\ref{Kanemele}\textbf{d}).   

\begin{figure*}[h!]
	\begin{tikzpicture}
	\node at (0,6) {\includegraphics[width=.45\textwidth]{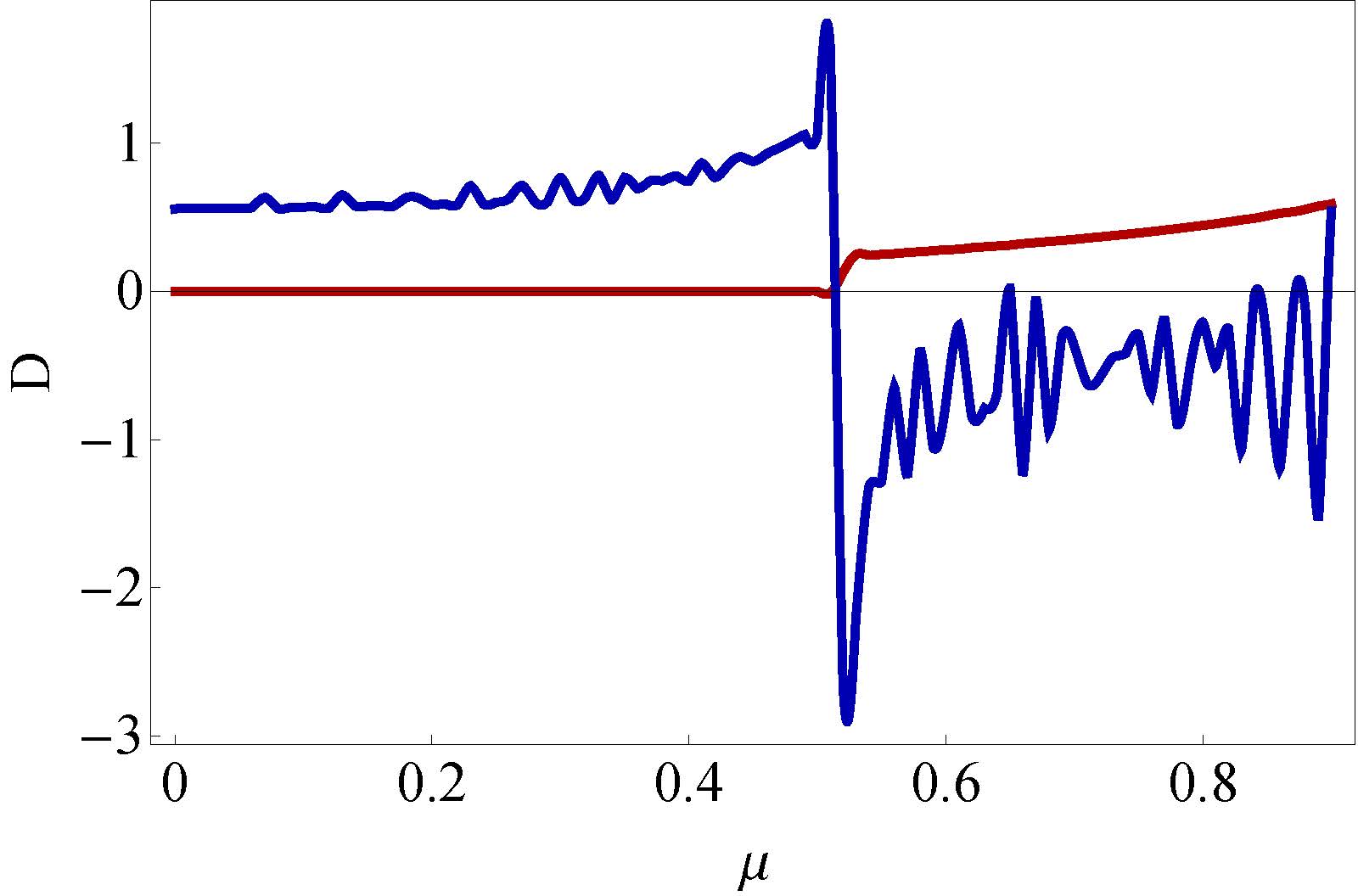}};
\node at (8,6)	{\includegraphics[width=.45\textwidth]{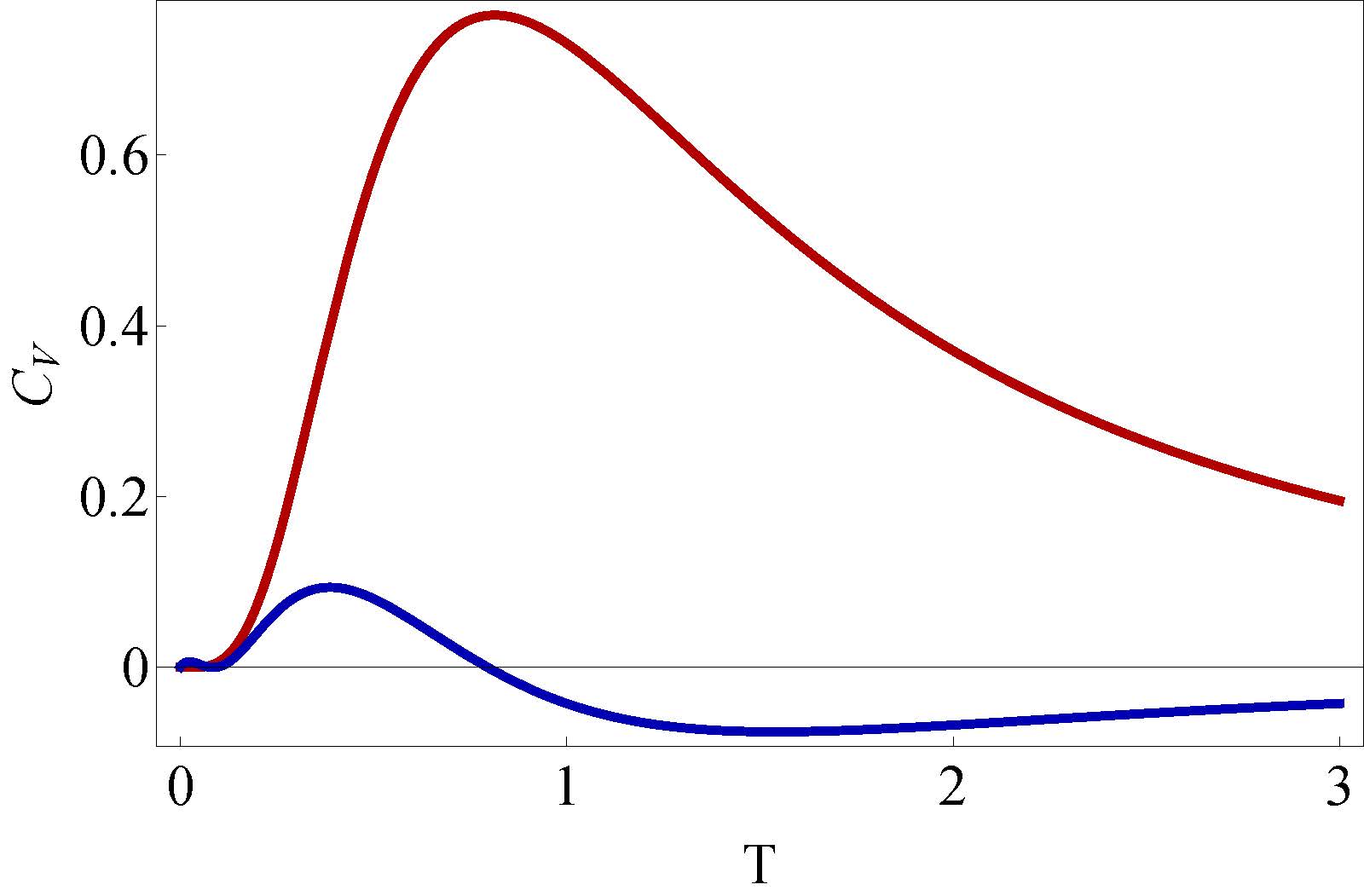}};
\node at (9.5,7.0)	{\includegraphics[width=.20\textwidth]{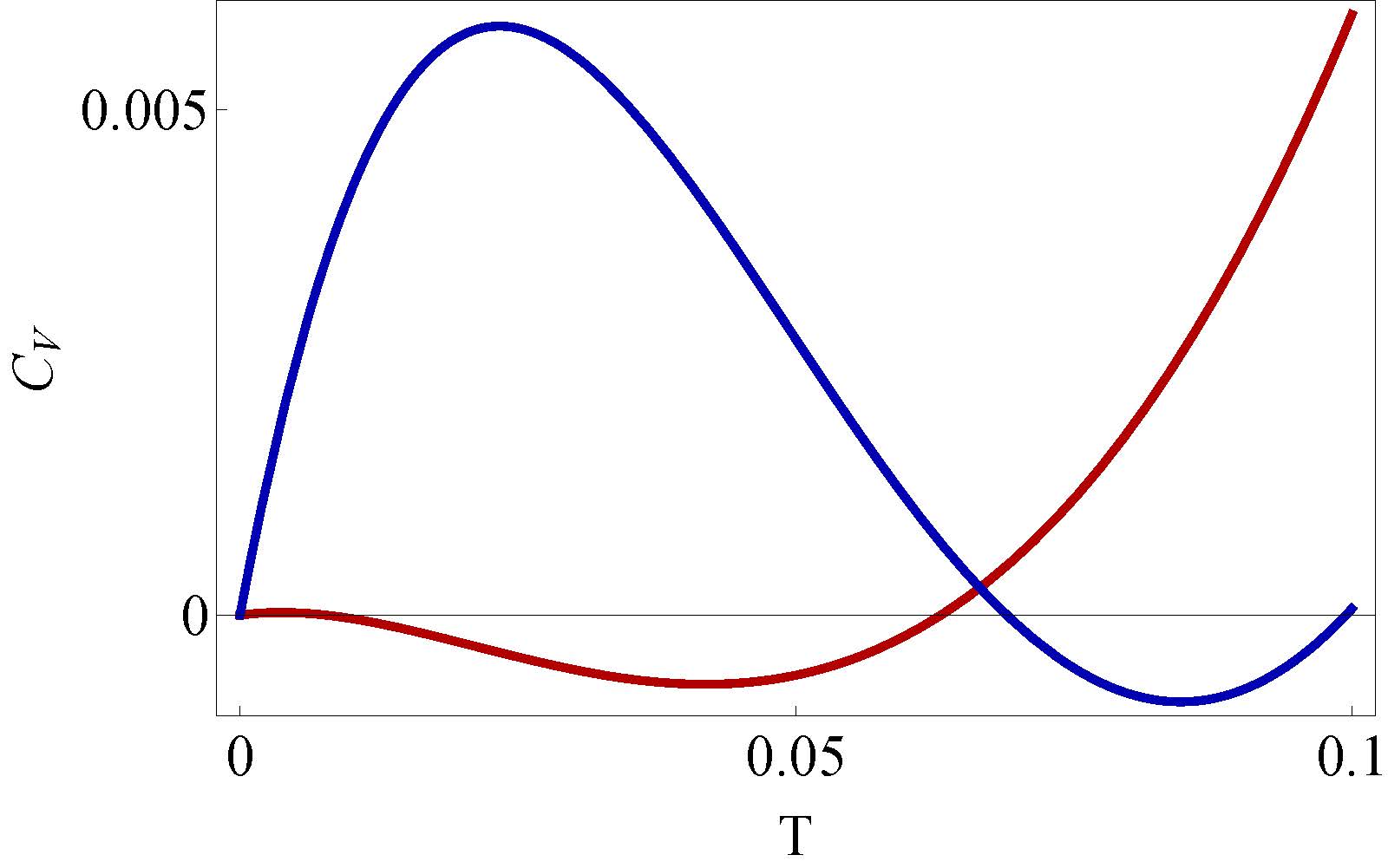}};
	\node at (-4,8.5) {\textrm{\textbf{a}}};
	\node at (4.4,8.5) {\textrm{\textbf{b}}};
\node at (-1,7.74) {\textrm{$\Phi_0$}};
\node at (-1,6.5) {\textrm{$\Phi_c$}};
\node at (7,7.74) {\textrm{$\Phi_c$}};
\node at (7,5.5) {\textrm{$\Phi_0$}};
	\end{tikzpicture}
	\caption{\textbf{Density of states $D_i=-\partial^2 \Phi_i/\partial \mu^2$  and heat capacity $C_{V,i}=-T \partial^2 \Phi_i/\partial T^2$ in the topological phase for $m=1$}. \textbf{a} At $T=0$, the density of states scales with the velocity of the electrons and is zero in the bulk (red) for $\mu<0.52$ (before the first conduction band), whereas at the boundary, the density of states should drop to zero after $\mu=0.52$ (blue). The noise for the boundary is due to numerical errors. \textbf{b} The heat capacity for the bulk (red) and the boundary (blue) for $\mu=0$. The inset shows the details of the low-$T$ behavior. The negative value of the heat capacity  should be interpreted as a lower total heat capacity for the whole system. The total heat capacity is always positive. }
	\label{heatdensityKM}
\end{figure*}
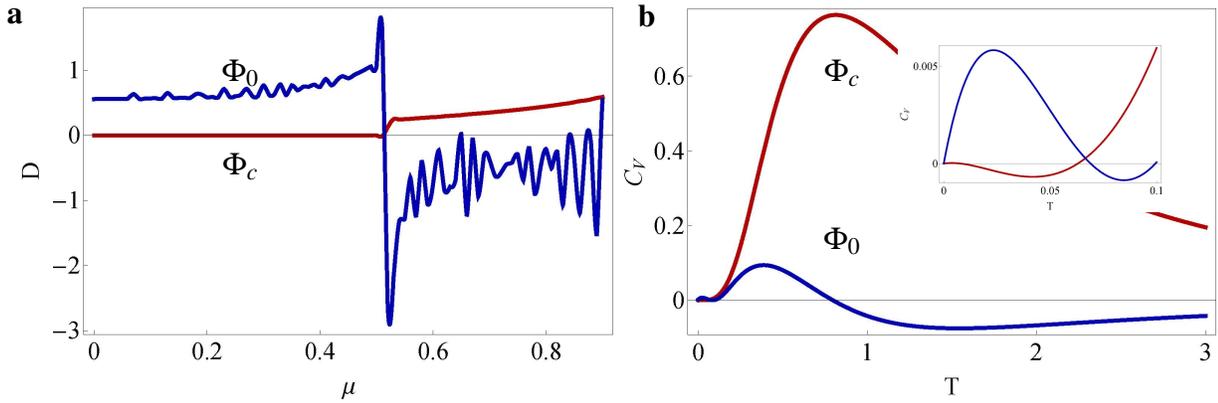

As an example and inspiration for experiments, we calculate the density of states $D_i=-\partial^2 \Phi_i/\partial \mu^2$ and the heat capacity $C_{V,i}=-T \partial^2 \Phi_i/\partial T^2$  in the topological phase (see Fig.~\ref{heatdensityKM}, where $\Phi_i$ denotes either $\Phi_C$ or $\Phi_0$ for the Kane-Mele model).
\subsection*{The BHZ model} The BHZ model is a theoretical model, first proposed to describe the appearance of a quantum spin Hall effect in mercury telluride-cadmium telluride quantum wells, upon tuning the thickness of the quantum well above a critical value~\cite{Bernevig}. This theoretical prediction was promptly followed by  the first experimental observation of this effect in 2007~\cite{Molenkamp}. Although the model was originally proposed to describe 2D systems, it has immediately raised the question whether an inversion of bulk bands could also arise in 3D materials, such as Bi$_x$Sb$_{1-x}$~\cite{Zhang}. The Hamiltonian to describe these 3D models in reciprocal space reads~\cite{Zhang}
\begin{equation}
H_\text{BHZ}(\textbf{k})= \epsilon_0(\textbf{k}) \mathbbm{1}_{4 \times 4} + 
\begin{pmatrix} 
M(\textbf{k})    & A_2 (k_x+i k_y) & 0              & A_1 k_z           \\ 
A_2 (k_x - ik_y) & -M(\textbf{k})  & A_1 k_z        & 0                 \\ 
0                & A_1 k_z         & M(\textbf{k})  & -A_2(k_x - i k_y) \\
A_1 k_z         & 0               & -A_2(k_x+ik_y) & -M(\textbf{k}) 
\end{pmatrix},
\end{equation}
where $\epsilon_0(\textbf{k})= C+ D_1 k_z^2 +D_2 (k_x^2+k_y^2)$, $M(\textbf{k})=M-B_1 k_z^2 - B_2 (k_x^2+k_y^2)$ and $M$, $A_1$, $A_2$, $B_1$, $B_2$, $C$, $D_1$ and $D_2$ are constant parameters in the model. In the numerical calculations, we set these parameters to one and rewrite the matrix into a tight-binding description. In Fig.~\ref{BHZ}\textbf{a-d}, we show the results for the thermodynamic potentials and their derivatives for the 3D BHZ model. The phase transition occurs for $M=0$ and is third-order at the edge and fourth-order in the bulk.
\begin{figure*}[h!]
	\begin{tikzpicture}
	\node at (8,6) {\includegraphics[width=.40\textwidth]{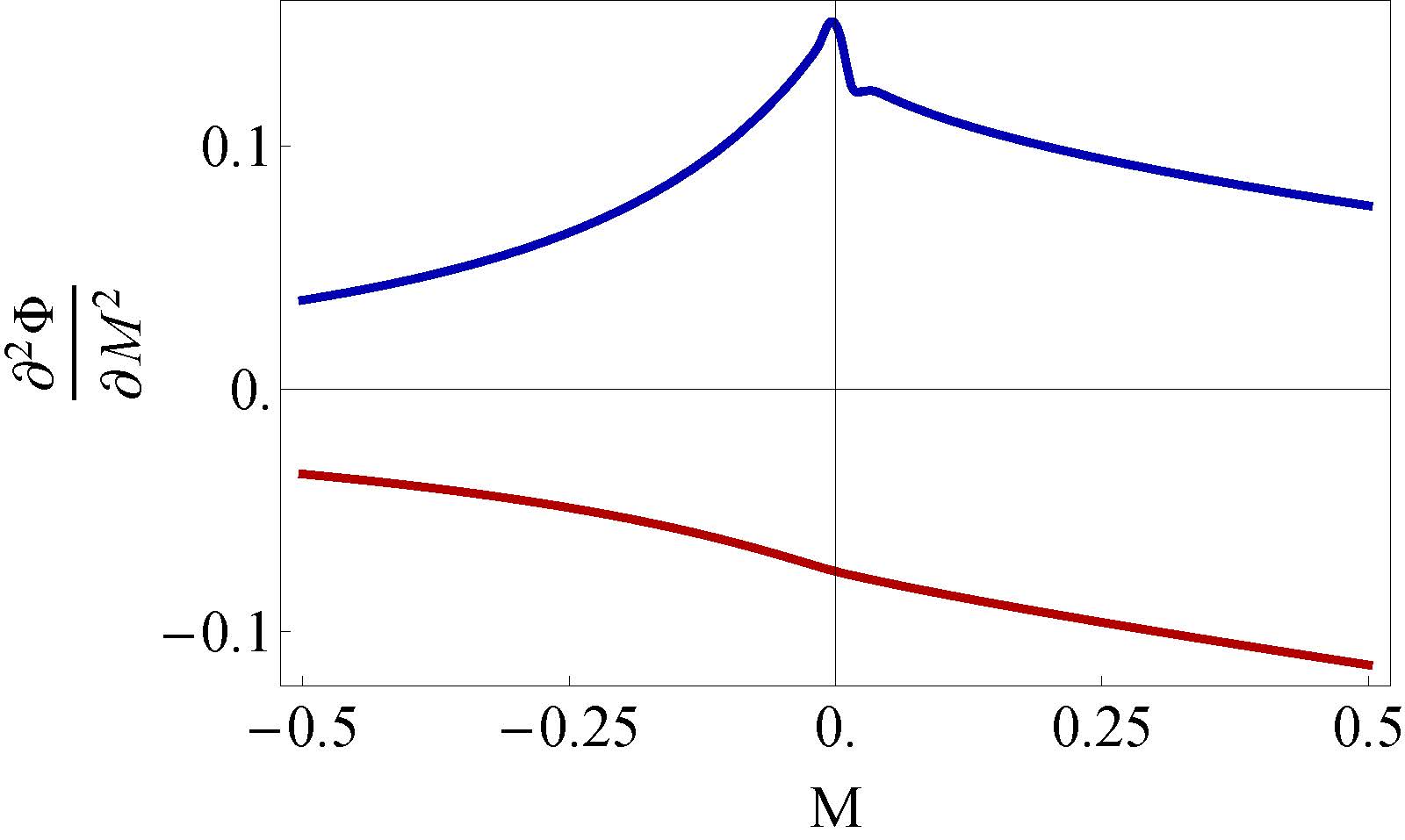}};
	\node at (8,1.5) {\includegraphics[width=.4\textwidth]{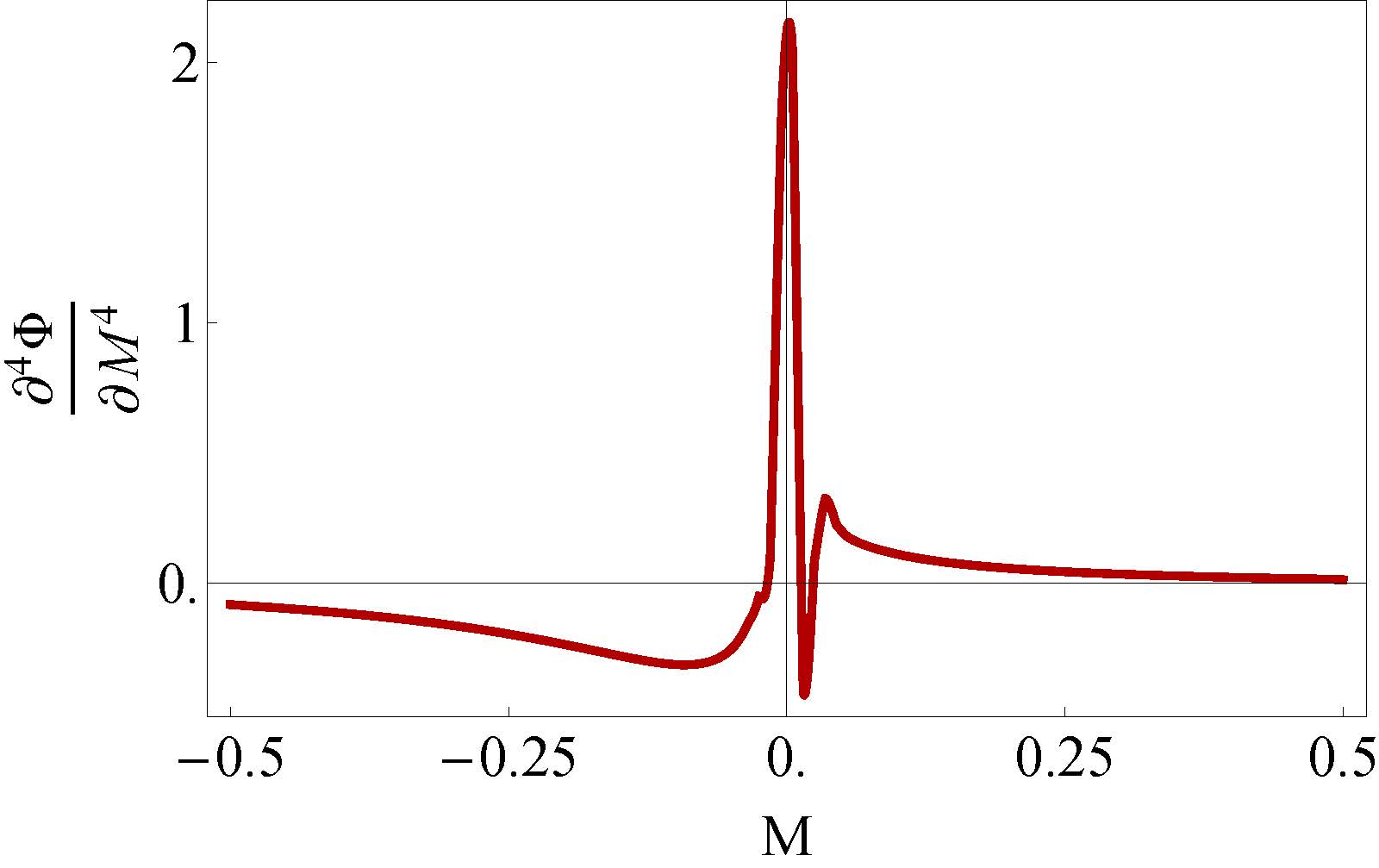}};
	\node at (0,1.5) {\includegraphics[width=.4\textwidth]{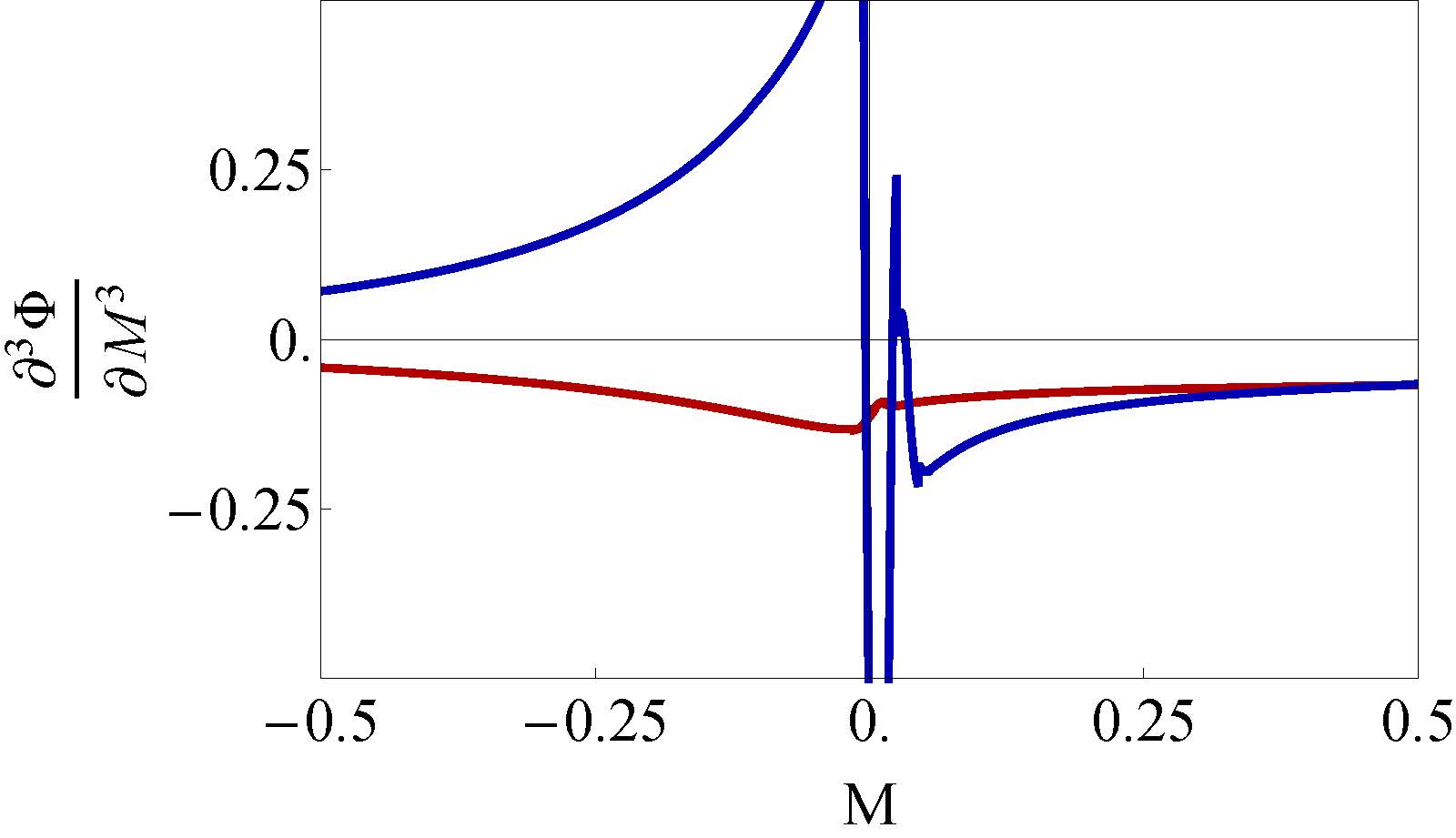}};
	\node at (0.3,6) {\includegraphics[width=.40\textwidth]{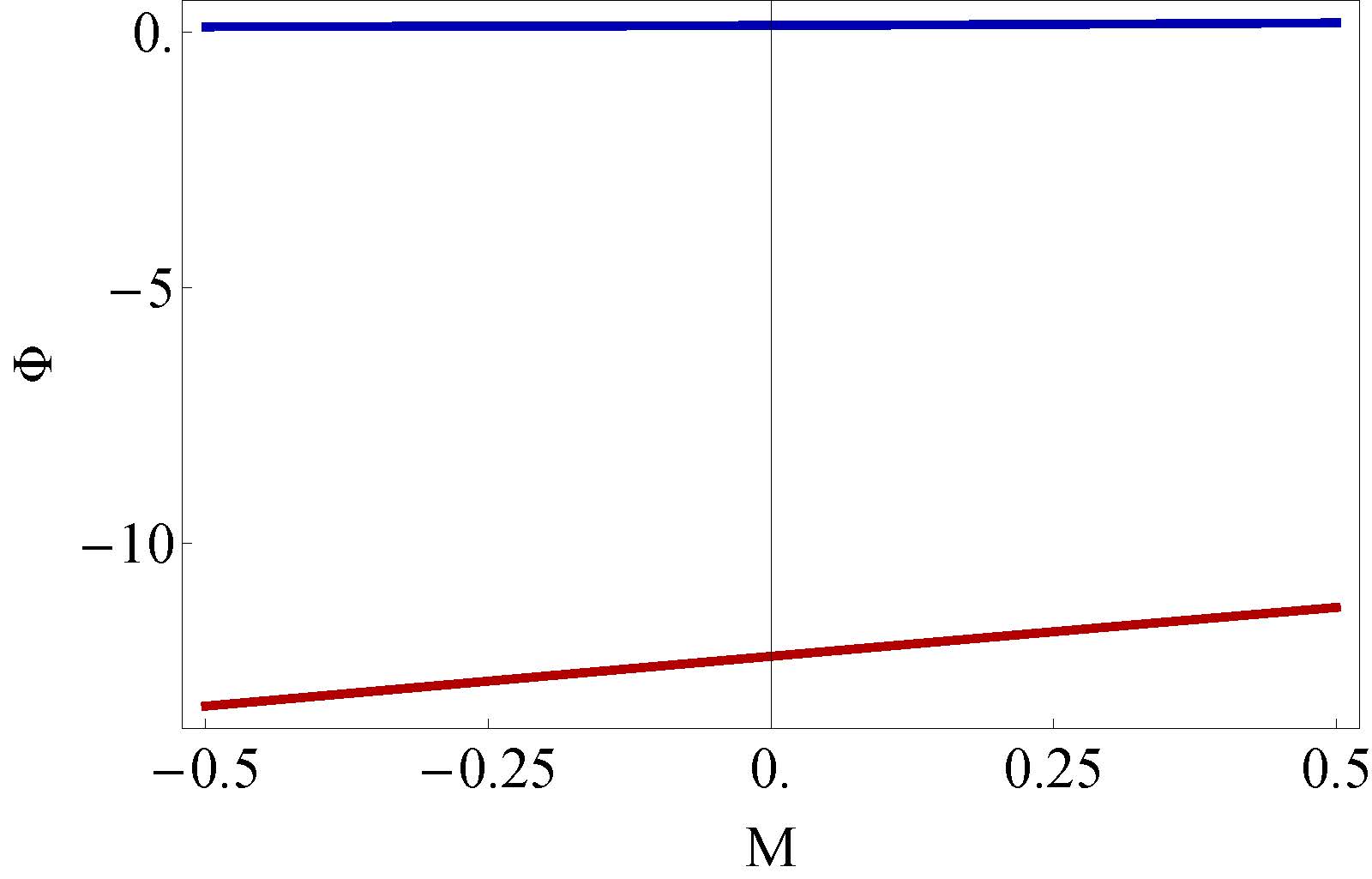}};
	\node at (-0.80,7.5) {\text{$\Phi_0$}};
	\node at (-0.80,5.2) {\text{$\Phi$}};
	\node at (-4.5,8) {\textrm{\textbf{a}}};
	\node at (4.4,8) {\textrm{\textbf{b}}};
	\node at (-4.5,3.5) {\textrm{\textbf{c}}};
	\node at (4.4,3.5) {\textrm{\textbf{d}}};
	\node at (2.8,6.4) {\includegraphics[width=0.08\textwidth]{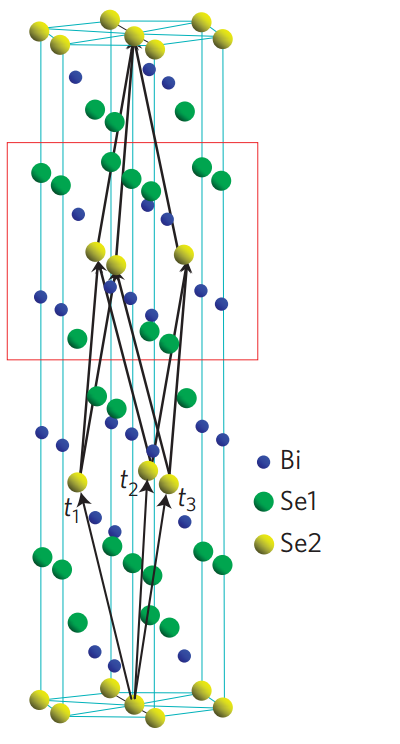}};
\node at (7.8,7.5) {\textrm{$\Phi_0''$}};
\node at (7.8,5.2) {\textrm{$\Phi_c''$}};
\node at (-0.80,2.7) {\textrm{$\Phi_0'''$}};
\node at (-0.80,1.10) {\textrm{$\Phi_c'''$}};
\node at (7.9,2.6) {\textrm{$\Phi_c^{(iv)}$}};
		\end{tikzpicture}
	\caption{\textbf{Behavior of the thermodynamic potentials and their derivatives for the BHZ model.} \textbf{a} The conventional potential $\Phi_c$ (red) and subdivision potential $\Phi_0$ (blue) and (\textbf{b, c, d}) its derivatives with respect to $M$  for the BHZ model with $A_1 = A_2 = B_1= B_2= C= D_1=D_2=1$. The phase change occurs at $M=0$. The subdivision potential shows a discontinuity in the third derivative, whereas the conventional potential shows a discontinuity in the fourth derivative. We should note that in \textbf{b} $\Phi_0''$ shows a spurious discontinuity near the phase transition (same for $\Phi_c'''$ in \textbf{c}). Therefore we considered the infinite system and verified that indeed a fourth-order phase transition occurs in the bulk (not shown).}
		\label{BHZ}
\end{figure*}
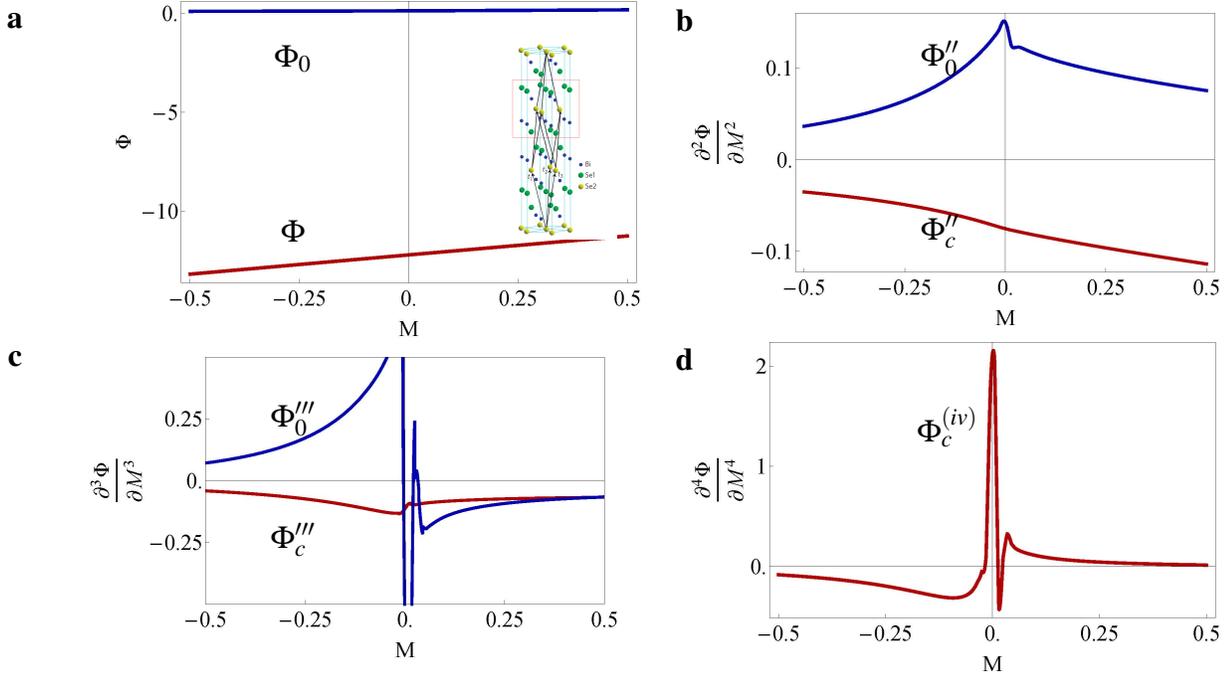 \newline

\section*{Conclusion}
Taken together, our results for 1D, 2D and 3D topological models reveal that the phase transition at the edge and bulk of the materials have a different order, and that this order is related to the spatial dimension $D$ of the quantum model: the edge phase transition has order $D$, whereas the bulk has order $D + 1$. Since our results are numerical, the next step would be to analytically examine topological systems within this framework and to verify the universalities that were shown here. It would also be interesting to find an interpretation of the critical exponents in the system, which are known for the Kitaev chain, since it is dual to the Ising chain~\cite{Greiter}. Furthermore, Hill thermodynamics might play a crucial role in understanding new types of topological phases such as finite-$T$ topological phases, interaction driven topological phases in the quantum valley Hall effect~\cite{MoraisSmith} and dissipation driven topological phases~\cite{Diehl}, due to its simplicity and because thermodynamics is a well-established theory. Finally, this thermodynamic description might trigger new experiments in the field, and the development of novel experimental tools, focusing on measurements of thermodynamic properties, such as the heat capacity and the density of states for 1D and 2D systems, in the bulk and at the boundary. 

\section*{Methods}
\label{Methods}
\section*{Hill thermodynamics}
The theory we use is based on the thermodynamics of finite-size systems developed by Terrell Hill in the 1960s~\cite{Hill}. Hill starts his analysis of small systems by considering an ensemble of $\mathcal{N}$ distinguishable, independent subsystems, all characterized by the volume $V$, temperature $T$ and chemical potential $\mu$. In the conventional thermodynamic limit of $\mathcal{N} \rightarrow \infty$, the whole ensemble is a macroscopic thermodynamic system, although each individual subsystem is not. The total energy of the system, besides the usual terms, also depends on $\mathcal{N}$, and therefore one can write for the internal energy  
\begin{equation}
\label{eq1}
dE_t= TdS_t - p \mathcal{N} dV +\mu dN_t + X d\mathcal{N},
\end{equation}
where $p$ is the mean pressure, $S$ is the entropy, $N$ is the number of particles within the system and the sub-index $t$ denotes properties of the whole ensemble. He also introduced the term $X = -\hat{p} V$, which can be regarded as another $pV$ work term that adds to the internal energy. Indeed, $X d \mathcal{N}$ corresponds to changing the volume of the system by increasing the number of subsystems with volume $V$, rather than changing $V$ per subsystem. By integrating equation (\ref{eq1}) in the grand canonical ensemble (keeping $\mu, V$ and $T$ constant), one finds
\begin{equation}
\label{eq2}
E_t=TS_t + \mu N_t - \hat{p} V \mathcal{N},
\end{equation}
and one identifies $X= -\hat{p} V = (E_t -TS_t -\mu N_t)/\mathcal{N}$. The next step is an important one, as it characterizes the essential mechanism of Hill thermodynamics. We define $\bar{E} = E_t/\mathcal{N}$ and $\bar{N}=N_t/\mathcal{N}$ as the appropriate thermodynamic ensemble averages of $E$ and $N$ per subsystem and $S=S_t/\mathcal{N}$ as the appropriate $S$ per subsystem. Notice that we did not put a bar over $S$ because it is a property of the complete probability distribution and not an average value in the same sense as $\bar{N}$ and $\bar{E}$. By using these definitions, we can rewrite equation (\ref{eq1})
\begin{eqnarray}
\label{eq3}
d\left(\mathcal{N}\bar{E} \right) &=& Td\left(\mathcal{N} S \right) - p \mathcal{N} dV +\mu d\left(\mathcal{N} \bar{N}\right) + X d\mathcal{N} \nonumber \\
\mathcal{N}d\bar{E}+ \bar{E} d\mathcal{N} &=& T\left(\mathcal{N} dS + S d\mathcal{N} \right) - p \mathcal{N} dV +\mu \left(\mathcal{N} d\bar{N} + \bar{N} d\mathcal{N} \right) + X d\mathcal{N} \nonumber \\
d\bar{E} &=& T dS - p dV + \mu d\bar{N},
\end{eqnarray}
where in the last line we substituted $X=\bar{E} - TS -\mu \bar{N}$ and we divided by $\mathcal{N}$ to obtain the differential equation for a single subsystem. Notice that the thermodynamic identity for a subsystem is still given by the usual thermodynamic expression. However, when comparing equation (\ref{eq3}) with equation (\ref{eq2}), we observe that upon integrating the expression in the grand canonical ensemble, the internal energy per system is 
\begin{equation}
\label{Hill}
\bar{E}=TS + \mu \bar{N} - \hat{p} V = TS - pV + \mu \bar{N} + \mathcal{E},
\end{equation}
where $\mathcal{E}:= (p-\hat{p})V$ defines the subdivision potential, which states that the internal energy is \textit{not} a linear homogeneous function of $S, V$ and $\bar{N}$. Equation (\ref{Hill}) is the main result of Hill\rq{s} analysis and captures the full non-extensive behavior of a finite-size subsystem in the grand-canonical ensemble. For a macroscopic system, the mean pressure and the integral pressure are equal $\hat{p}=p$, but this does not hold in general for a finite-size system, where doubling the volume $V$ will not double the internal energy because of the finite size and edge effects. As a result, the subdivision potential vanishes in the thermodynamic limit and we obtain the usual expressions for the macroscopic system.

\section*{Gibbs thermodynamics}
We should note that, although the procedure looks trivial, it differs in an important way from the conventional approach developed by Gibbs when considering boundaries in a system. In Gibbs\rq{} procedure, one can split the contribution from the bulk ($B$) and the boundary ($b$), such that the internal energy of the bulk is
\begin{equation}
dE_B= dS_B T_B -p dV + \mu_B dN_B,
\end{equation}
and the energy for the boundary is 
\begin{equation}
dE_b= dS_b T_b - \gamma dA + \mu_b dN_b.
\end{equation}
In equilibrium, $T_B=T_b=T$, $\mu_B=\mu_b=\mu$, and the total energy of the system reads 
\begin{equation}
dE_{B+b}= dE_B +dE_b =  T d\left(S_B+S_b \right) - pdV - \gamma dA + \mu d\left(N_B+N_b \right),
\end{equation}
where we encounter and additional $-\gamma dA$ term in the thermodynamic identity. This is a crucial difference when comparing Gibbs' and Hill's approaches. Whereas Gibbs assumes that the boundary contributes in the above way, Hill connects the boundary to the bulk in a natural way and finds that the subdivision potential might correspond to $\gamma A$ and additional terms. Therefore, Gibbs' procedure works perfectly well when the bulk and the boundary can be considered independent, or when an effective theory can be written down for both parts. However, this theory is not useful when considering a system with a boundary phase transition, in which the boundary merges with the bulk, as it is the case for topological phase transitions. \newline

In Ref.~\onlinecite{Quelle}, it was shown that we can use Hill's formalism to describe topological models, and that the subdivision potential provides a thermodynamic description of the edge states. Here, we recall the main steps for completeness. To connect Hill thermodynamics to topological models, one considers a system of finite size $V$ in contact with an environment characterized by $T$ and $\mu$. The grand potential is
\begin{equation}
\label{thermo}
\Phi= \bar{E}-TS-\mu \bar{N}  = \Phi_c + \Phi_0,
\end{equation}
with $\Phi_c= -pV$ and $\Phi_0=\mathcal{E}$ the subdivision potential. One of the strengths of Hill's approach is that the thermodynamic identities are kept the same as in conventional thermodynamics. The grand potential of this system in the framework of statistical mechanics reads
\begin{equation}
\label{statistical}
\Phi= -k_{\text{B}}T \log \left\{ \textrm{Tr} \exp{ \left[ -\left(H-\mu N \right) / (k_{\text{B}}T) \right] }   \right\},
\end{equation}
where $k_{\text{B}}$ denotes the Boltzmann constant and $H$ the Hamiltonian of the topological insulator. Now, by comparing equations (\ref{thermo}) and (\ref{statistical}), one finds an expression for the subdivision potential. In principle, $\Phi_0$ can take any form, since it does not scale with the volume. However, when the system is large, we can use the Ansatz proposed in Ref.~\onlinecite{Quelle}
\begin{equation}
\label{thermo2}
\Phi(\mu,T,V) = \Phi_c(\mu,T) V + \Phi_0(\mu,T) V/W,
\end{equation}
where $V$ is the $D$ dimensional volume and $W$ is the distance between the boundaries, such that the subdivision potential $\Phi_0$ scales with the boundary of the system and captures its topological properties~\cite{Quelle}. By calculating the eigenvalues of the Hamiltonian, a linear fit is used to find the correspondence between equations (\ref{statistical}) and (\ref{thermo2}), and to obtain the topological behavior in a thermodynamic description. 
\section*{Numerics}
Two independent calculations were performed to obtain the grand potential for the Kitaev chain, the SSH model and the BHZ model (3D), one in Mathematica and one in C++, and were shown to be exactly the same. In C++, the Eigen package was used to calculate the eigenvalues of the matrices. Due to the unstable behavior of the free energy near the phase transition, we needed to consider larger system sizes in this region to obtain an appropriate linear fit (i.e. a linear fit without systematic errors from finite-size effects). The parameter values for each system are described below (in natural units). All the data is available upon request.\\
\textbf{In the Kitaev chain}, the system sizes vary between $200 \leq L \leq 250$ for $0.3 \leq \mu \leq 0.45$ and $ 0.55 \leq \mu \leq 0.7$, between $1000 \leq L \leq 1050$ for $0.45 \leq \mu \leq 0.49$ and $ 0.51 \leq \mu \leq 0.55$ and $4000 \leq L \leq 4050$ for $0.49 \leq \mu \leq 0.51$. The datapoints are obtained via steps $\delta \mu =0.001$. In the Kitaev ring, all the lengths are $1000 \leq L \leq 1050$ with $\delta \mu = 0.002$. The entropy is obtained with the same parameters for a range of values for $T$. For calculating the heat capacity, we used $L=4000$ and varied $T$ in steps of $\delta T=0.0001$. The results for the finite-$T$ phase diagram were obtained for a range of values for $t < 1.2$ with $\delta t=0.2$ at different $T$ in steps of $\delta T=0.05$, all between length  $200 \leq L \leq 250$ for $0 \leq 2 \mu \leq t $ and $1000 \leq L \leq 1050$ otherwise. \\
\textbf{In the SSH model}, the system sizes vary between $200 \leq L \leq 250$ for $-0.5 \leq \Delta \leq -0.034$ and $0.006\leq \Delta \leq 0.50$, between  $500 \leq L \leq 550$  for $-0.034 \leq \Delta \leq-0.01$ and $0.006\leq \Delta \leq 0.022$ and $1000 \leq L \leq 1050$ otherwise. The datapoints are obtained in steps of $\delta \Delta =0.002$. The results for the finite-$T$ phase diagram were obtained for a range of values for $t_1, t_2 < 1.2$ at different $T$ in steps of $\delta T =0.05$, all between lengths $200 \leq L \leq 250$.\\
\textbf{In the Kane-Mele model}, the system sizes vary between $600 \leq L \leq 610$ with step size $\delta \mu=0.001$ for the free energy. The results for the density of states is obtained for $400 \leq L \leq 450$ and the heat capacity for $400 \leq L \leq 450$ and in steps of $\delta T=0.05$.\\
\textbf{In the BHZ model}, the system size $50 \leq L \leq 60$ for $-0.5 \leq M < -0.15$ and $ 0.25 < M \leq 0.5$ and $200 \leq L \leq 210$ is smaller than before because these calculations involve large matrices. To obtain the total energy, we summed over the wave vectors $k_x$ and $k_y$ for $2 \pi n/n_{max}$, where $0 < n \leq n_{max}=100$. Furthermore, all parameters, except M, are set to 1, and we used $\delta M = 0.01$ as step-size between datapoints. 

\clearpage



\section*{Acknowledgements}
We would like to thank Giandomenico Palumbo and Dirk Schuricht for useful discussions. This work is part of the D-ITP consortium, a program of the Netherlands Organization for Scientific Research (NWO) that is funded by the Dutch Ministry of Education, Culture and Science (OCW).

\section*{Author contributions statement}
All authors contributed in the analysis of the data, A. Quelle and S.N. Kempkes prepared the calculations of all figures, and C. Morais Smith and S.N. Kempkes wrote the main manuscript text. 

\section*{Additional information}
Correspondence and requests for materials should be addressed to C. Morais Smith~(email: c.moraissmith@uu.nl). The authors declare that they have no competing financial interests. 

\end{document}